\documentclass[12pt]{iopart}
\usepackage{epsfig}

\newcommand{\be}{\begin{equation}}
\newcommand{\ee}{\end{equation}}
\newcommand{\bea}{\begin{eqnarray}}
\newcommand{\eea}{\end{eqnarray}}
\newcommand{\beastar}{\begin{eqnarray*}}
\newcommand{\eeastar}{\end{eqnarray*}}
\newcommand{\nn}{\nonumber\\}

\newcommand{\eq}[1]{~(\ref{#1})}

\newcommand{\order}{{{\mathcal O}}}

\newcommand{\ar}{\leftarrow}

\newcommand{\tw}{t_{\rm w}}
\newcommand{\mw}{m_{\rm w}}
\newcommand{\Ew}{E_{\rm w}}

\newcommand{\dt}{\Delta t}
\newcommand{\dd}[1]{\frac{\partial}{\partial#1}}
\newcommand{\Tg}{T_{\rm g}}
\newcommand{\Teff}{T_{\rm eff}}
\newcommand{\BM}{Barrat and M\'ezard}
\newcommand{\JB}{Junier and Bertin}

\newcommand{\D}{\Delta}

\renewcommand{\k}{\kappa}
\newcommand{\mb}{\mu}
\newcommand{\Psc}{{\mathcal{P}}}
\newcommand{\Msc}{{\mathcal{M}}}
\newcommand{\mbsc}{{\mathcal{U}}}
\newcommand{\om}{\omega}
\newcommand{\ta}{\tau}
\newcommand{\Csc}{{\mathcal{C}}}
\newcommand{\Rsc}{{\mathcal{R}}}
\newcommand{\Xsc}{{\mathcal{X}}}
\newcommand{\Q}{{\mathcal{Q}}}
\newcommand{\rh}{\gamma}
\newcommand{\lam}{\lambda}

\newcommand{\cT}{c}
\newcommand{\cM}{{c^{\rm M}}}
\newcommand{\dT}{\eta}
\newcommand{\dM}{\eta^{\rm M}}

\newcommand{\cTT}[1]{c_{#1}}

\newcommand{\cMM}[1]{c^{\rm M}_{#1}}

\begin{document}

\title{Trap models with slowly decorrelating observables}

\author{Peter Sollich\footnote[1]{Email peter.sollich@kcl.ac.uk}}

\address{King's College London, Department of Mathematics, London WC2R
2LS, UK}

\begin{abstract}
We study the correlation and response dynamics of trap models of
glassy dynamics, considering observables that only partially
decorrelate with every jump. This is inspired by recent work on a
microscopic realization of such models, which found strikingly simple
linear out-of-equilibrium fluctuation-dissipation relations in the
limit of slow decorrelation. For the \BM\ model with its entropic
barriers we obtain exact results at zero temperature $T$ for arbitrary
decorrelation factor $\k$. These are then extended to nonzero $T$,
where the qualitative scaling behaviour and all scaling exponents can
still be found analytically.  Unexpectedly, the choice of transition
rates (Glauber versus Metropolis) affects not just prefactors but also
some exponents. In the limit of slow decorrelation even complete
scaling functions are accessible in closed form. The results show that
slowly decorrelating observables detect persistently slow
out-of-equilibrium dynamics, as opposed to intermittent behaviour
punctuated by excursions into fast, effectively equilibrated states.
\end{abstract}




\section{Introduction}

Trap
models~\cite{Bouchaud92,BarMez95,Dyre87,BouDea95,BouComMon95,MonBou96,%
Head00,BerBou02,FieSol02,Ritort03,Sollich03}
have been recognized in recent years as powerful models of glassy,
non-equilibrium dynamics. They describe the motion of a system through
its phase space, simplified to a picture of thermally activated
hopping in a landscape of traps of energy $E$. The simplest case is
that of mean-field trap models, where all traps are taken as mutually
accessible and 
the rate for a transition between two traps depends only on their
energies. An alternative motivation for considering trap models is
provided by the problem of diffusion in disordered media; here the
traps are located in a physical space of low dimension (say, $d=1,2$
or 3) and the spatial organization of the traps has to be accounted
for~\cite{HauKeh87,RinMaaBou00,RinMaaBou01,BerBou03}.

Our motivation in this paper arises from a recent interesting study of
a microscopic realization of trap models on the basis of the number
partitioning problem~\cite{JunKur04,JunBer04}. Each partition of $N$
numbers between two piles can be associated with the state of a system
of $N$ Ising spins $s_i=\pm 1$, and an energy function is then defined
to measure how far from optimal the partition is. The dynamics
considered by \JB\ in~\cite{JunBer04} is that at
every step $K$ spins are selected and have their values randomized;
the new state is then accepted with a probability given by a
Metropolis acceptance factor. \JB\ argue that this system evolves
in a manner analogous to the trap model considered by
\BM~\cite{BarMez95}: a new state or trap is selected essentially at
random at every step, and accepted according to the Metropolis (or
Glauber, see below) probability. To justify the assumption that each
new state is effectively random one requires $K\gg 1$; for finite $K$
there is a crossover time, exponentially large in $K$, beyond which
this simplification no longer applies~\cite{JunBer04}. In contrast to the usual
assumption made in trap models, however, an observable such as the
(randomly staggered) magnetization does not decorrelate fully with
every transition between states. Instead, it decorrelates by a factor
$\k=1-K/N$, since only $K$ out of the $N$ spins are updated. \JB\
showed that the consequences of this are rather profound, and most
interesting in the limit of slow decorrelation $\k\to 1$ ($K\ll
N$). In particular, they found that fluctuation-dissipation (FD)
relations between response and correlation functions had the simple
straight-line form expected for systems with well-defined (effective)
temperatures~\cite{CugKur93,CugKur94,CugKurPel97,CriRit02b}. The
corresponding temperature was equal to the bath temperature $T$ down
to {\em half} the glass transition temperature $\Tg$; for lower $T$,
it remained pinned to $\Teff=\Tg/2$. This unusual transition between
apparent equilibrium and non-equilibrium dynamics {\em within} the
glass phase was interpreted as due to a change from activated to
entropic slowing down.

Our aim in this paper is to complement the work of~\cite{JunBer04},
which used mainly numerical simulation and simple scaling
estimates, with an analytical study of the dynamics of slowly
decorrelating observables in trap models. This will allow us, for example,
to verify and refine estimates of scaling exponents
in~\cite{JunBer04}, but also yield more detailed insights into the
nature of the dynamics. In our approach it is also a simple
matter to compare different choices of the transition rates
(Metropolis and Glauber) and we will see that this has some unexpected
consequences, with effects not just on prefactors but also scaling
exponents.

To allow a direct comparison with the work of~\cite{JunBer04} we focus
mainly on the \BM\ model itself; the extension of this to slowly
decorrelating observables is defined in
Sec.~\ref{sec:definitions}. General expressions for correlation and
response functions are then derived in Sec.~\ref{sec:corr_resp}, and
simplified in Sec.~\ref{sec:long-time} for the interesting scaling regime of
long times. As a prelude to the analysis of slowly decorrelating
observables proper, we consider in Sec.~\ref{sec:kappa0} first the
standard, fully decorrelating case $\k=0$ at nonzero temperature,
extending previous analytical results for the FD behaviour in the
limit $T\to 0$~\cite{Sollich03}. Sec.~\ref{sec:T0} then looks at
arbitrary decorrelation factors $\k$, but first at zero temperature,
where closed form results can still be obtained. The most
general case of $\k>0$ and $T>0$ is then analysed in
Sec.~\ref{sec:kappa_T}, where we also provide results from numerical
solutions of the integral equations for the scaling functions. In
Sec.~\ref{sec:bouchaud}, finally, we extend the analysis to more general
trap models with a multiplicative dependence of the transition rates
on the applied field~\cite{Ritort03}; this covers in particular
Bouchaud's original trap model~\cite{Bouchaud92} with its purely
activated dynamics. Sec.~\ref{sec:conclusion} summarizes the results and
discusses the dynamics of slowly decorrelating observables in a wider
context.

\section{Model definition}
\label{sec:definitions}

Motivated by the arguments discussed in the introduction, we consider
a modified \BM\ model~\cite{BarMez95} where the magnetizations before
and after a jump between two traps are at least partially
correlated. Transitions can still take place between arbitrary trap
{\em energies}, but are more likely between traps with magnetizations that are
sufficiently close to each other.

The dynamics of the system is described by the probability $P(E,m,t)$
that the system is in a trap with energy $E$ and magnetization $m$ at
time $t$.  Of course $m$ is in principle a generic observable, but we
continue to use the term magnetization for definiteness. We assume
that, starting from a trap with energy $E'$ and magnetization $m'$, a
jump to a new trap with energy $E$ and magnetization $m$ is attempted
with probability $\rho(E)\rho(m|m')$. The first factor is simply the
density of states of trap energies, reflecting the assumption that
transitions between arbitrary energy levels are possible. The factor
$\rho(m|m')$, on the other hand, can incorporate various degrees of
correlation between $m'$ and $m$; if $m$ is typically close to $m'$,
then decorrelation of $m$ over time is slow. The factorization
$\rho(E)\rho(m|m')$ 
assumes that $E$ and $m$ are selected independently, which in a spin
system is a reasonable assumption if $m$ is some staggered
magnetization that is uncorrelated with the energy. In a
continuous-time description, the time evolution of $P(E,m,t)$ is then
governed by the master equation
\bea
\fl\dd{t} P(E,m,t) &=& -\Gamma(E,m)P(E,m,t) + 
\nn
& & {}+{} 
\int\! dE'\,dm'\, \rho(E) \rho(m|m') w(E-E'-hm+hm') P(E',m',t)
\label{master2}
\eea
Here $w(\D E)$ is the probability with which a proposed transition is
accepted. In writing the energy change $\D E=E-hm-(E'-hm')$ on which
this depends we have allowed for the presence of a field $h$
thermodynamically conjugate to $m$, in order to be able to deduce
response properties of $m$. We consider primarily the Glauber form
$w(\D E)=1/[1+\exp(\beta\D E)]$ of the acceptance probability
that was used in previous theoretical
studies~\cite{BarMez95,Sollich03,Bertin03}, but also compare with the
Metropolis choice, $w(\D E)=1$ for $\D E<0$ and $w(\D E)=\exp(-\beta\D E)$
for $\D E\geq 0$, which is common in
simulations~\cite{JunBer04,Bertin03}. Here and throughout
$\beta=1/T$ denotes the inverse temperature. In the first
term on the right of\eq{master2}, finally, we have defined
\be
\Gamma(E,m) = \int\! dE'\,dm'\, \rho(E') \rho(m'|m) w(E'-E-hm'+hm)
\label{GamEm}
\ee
which can be thought of as the total rate of leaving a trap with
energy $E$ and magnetization $m$. 

Our dynamics should be capable of describing thermodynamic equilibrium
at high temperatures, and thus obey detailed balance. In line with our
assumption of independence of energies and magnetizations in an
attempted jump, let us assume the overall density of states also
factorizes into $\rho(E)\rho(m)$. It is then easy to check that the
dynamics\eq{master2} obeys detailed balance as long as
$\rho(m'|m)\rho(m)$ is symmetric under interchange of $m'$ and $m$.
To be specific, we take $\rho(m)$ to be a Gaussian with zero mean and
unit variance, and $\rho(m'|m)$ a Gaussian with mean $\k m$ and
variance $1-\k^2$. (Since we will only need the second order
statistics of $m$ and $m'$, this assumption in fact constitutes no
loss of generality.) The original \BM\ model then corresponds to
Glauber dynamics with $\k=0$, while in the opposite limit $\k=1$ the
magnetization 
remains frozen to its initial value. For intermediate values, $m$
decorrelates by a factor $\k$ with each jump, and we expect the
interesting behaviour seen by \JB~\cite{JunBer04} to occur in the
limit of slow decorrelation $\k\to 1$ (but keeping $\k<1$).

\section{General expressions for correlation and response}
\label{sec:corr_resp}

For the calculation of the correlation function of $m$, the field can
be set to zero. The master equation\eq{master2} then simplifies to
\be
\fl\dd{t} P(E,m,t) = -\Gamma(E)P(E,m,t) + 
\int\! dE'\,dm'\, \rho(E) \rho(m|m') w(E-E') P(E',m',t)
\label{master_h0}
\ee
with
\be
\Gamma(E) = \int\! dE'\, \rho(E') w(E'-E)
\label{GamEm_h0}
\ee
The propagator $P(E,m,t-\tw|\Ew,\mw)$ obeys this equation with the
initial condition $P(E,m,0|\Ew,\mw)=\delta(E-\Ew)\delta(m-\mw)$ and
gives the correlation function as
\be
\fl C(t,\tw) = \int\!
dE\,dm\,d\Ew\,d\mw\,m\,\mw\,P(E,m,t-\tw|\Ew,\mw) \rho(\mw)P(\Ew,\tw)
\ee
Here we have used that, for $h=0$, $P(\Ew,\mw,\tw) =
\rho(\mw)P(\Ew,\tw)$; this is true as long as the initial condition
has the same structure, e.g.\ for a quench from thermal equilibrium at
high temperature. From\eq{master_h0}, the function
$\mb(E,t-\tw|\Ew,\mw) = \int\!dm\,m\,P(E,m,t-\tw|\Ew,\mw)$ obeys,
using $\int\!dm\,m\rho(m|m') = \k m'$,
\be
\fl\dd{t} \mb(E,t|\Ew,\mw) = -\Gamma(E)\mb(E,t|\Ew,\mw) + 
\k \int\! dE'\, \rho(E) w(E-E') \mb(E',t|\Ew,\mw)
\label{mu_general}
\ee
Because this equation is linear, the factor $\mw$ from the initial
condition $\mb(E,0|\Ew,\mw)=\mw\delta(E-\Ew)$ pulls through and we can
write $\mb(E,t|\Ew,\mw) = \mw \mb(E,t|\Ew,1) \equiv \mw \mb(E,t|\Ew)$,
dropping the constant argument 1. The reduced magnetization
$\mb(E,t|\Ew)$ then obeys
\be
\fl \dd{t} \mb(E,t|\Ew) = -\Gamma(E)\mb(E,t|\Ew) + 
\k \int\! dE'\, \rho(E) w(E-E') \mb(E',t|\Ew)
\label{mb_forward}
\ee
with $\mb(E,0|\Ew)=\delta(E-\Ew)$. For our purposes more useful,
however, is the corresponding backward equation,
\be
\fl \dd{t} \mb(E,t|\Ew) = -\Gamma(\Ew)\mb(E,t|\Ew) + 
\k \int\! dE'\, \mb(E,t|E') \rho(E') w(E'-\Ew) 
\label{mb}
\ee
This is because the correlation function can be written as
\be
C(t,\tw) = \int\! d\Ew\, M(t-\tw|\Ew)P(\Ew,\tw)
\label{corr}
\ee
with
\be
M(t-\tw|\Ew) = \int\! dE\,\mb(E,t-\tw|\Ew)
\label{Mdef}
\ee
By integrating\eq{mb} we see that this obeys
\be
\dd{t} M(t|\Ew) = -\Gamma(\Ew)M(t|\Ew) + 
\k \int\! dE'\, M(t|E') \rho(E') w(E'-\Ew)
\label{M_eq}
\ee
with $M(0|\Ew)=1$; the forward equation\eq{mb_forward}, on the other
hand, would yield an expression for $\partial M/\partial t$ which
still involves $\mb(E,t|\Ew)$. The physical meaning of $M(t-\tw|\Ew)$
is as follows: if we start in a state with energy $\Ew$ and
magnetization $\mw$, then $\mw M(t-\tw|\Ew)$ is the average
magnetization a time $t-\tw$ later. 

For the case $\k=0$ the solution of\eq{mb} and\eq{M_eq} is trivial,
\bea
\mb(E,t|\Ew) &=& \delta(E-\Ew)\exp(-\Gamma(\Ew)t)
\\
M(t|\Ew)&=&\exp(-\Gamma(\Ew)t)
\label{kappa0}
\eea
and one retrieves the standard result for the hopping correlation
function~\cite{BarMez95,Ritort03,Sollich03,Bertin03}. For $\k>0$, the
Laplace transform (LT) of\eq{mb} can be easier to work with; with $s$
conjugate to $t$, this reads
\be
\fl [s+\Gamma(E)]\hat{\mb}(E,s|\Ew) - \delta(E-\Ew) = 
\k \int\! dE'\, \rho(E) w(E-E') \hat{\mb}(E',s|\Ew)
\label{mb_LT}
\ee

By the same reasoning that lead to\eq{corr}, the response to a field
impulse of amplitude $h$ and duration $\dt$, applied at time $\tw$,
can be written as
\be
h\dt\, R(t,\tw) = \int\! dE\,dm\, m\,M(t-\tw|E)P(E,m,\tw+\dt)
\label{R_start}
\ee
The change in $P(E,m,\tw+\dt)$ from its value without the field (the
latter does not contribute to $R$ because it is symmetric in $m$) is
\bea
\fl \D P(E,m,\tw+\dt) &=& \dt\,\int\!d\Ew\,d\mw \times
\label{dP}\\
\fl & & \left[
 \rho(E)  \rho(m |\mw) \D w(E -\Ew-hm +h\mw) \rho(\mw) P(\Ew, \tw)
\right.
\nonumber\\
& &
\left.
-\rho(\Ew) \rho(\mw|m)  \D w(\Ew-E -h\mw+hm)  \rho(m)  P(E,  \tw)
\right]
\eea
where $\D w(E-\Ew-hm+h\mw)=w(E-\Ew-hm+h\mw)-w(E-\Ew)$ is the change of
the acceptance probability caused by the field. Expanding this to
linear order in $h$ as $\D w(E-\Ew-hm+h\mw)=h(\mw-m)w'(E-\Ew)$
and carrying out the integrations over $m$ and $\mw$ gives
\bea
\fl R(t,\tw) &=& (1-\k) 
\int\! dE\,d\Ew\,
M(t-\tw|E)\times
\nonumber\\
\fl & &\times 
\left[-w'(E-\Ew)\rho(E)P(\Ew,\tw)-w'(\Ew-E)\rho(\Ew)P(E,\tw) \right]
\label{resp}
\eea
Note that $w'(\cdot)$ is negative since the acceptance probability
$w(\cdot)$ decreases with increasing energy change; thus both terms in
the expression for the response are positive. Explicitly, we have for
the Glauber case $-w'(\D E)=\beta\exp(\beta \D E)/[1+\exp(\beta \D
E)]^2$ and for Metropolis $-w'(\D E)=\Theta(\D E)\beta\exp(-\beta \D
E)$, with $\Theta(\cdot)$ the usual Heaviside step function.

Summarizing, to calculate the correlation and response we need
$P(E,\tw)$, i.e.\ the solution of the original \BM\ model without a
field, and $M(t-\tw|\Ew)$ from\eq{M_eq}. The dependence on $\k$ is
only through the latter.

\section{Long-time scaling}
\label{sec:long-time}

From now on we consider mostly an exponential density of trap
energies, $\rho(E)=\exp(E)$ for $E<0$; the glass transition
temperature is then $\Tg=1$. We will also focus on the long-time
scaling limit where memory of the initial conditions is lost and
typical trap depths are large, $|E|\gg 1$. The exit rate from such
deep traps is
\be
\fl
\Gamma(E) = \int_{-\infty}^0\! dE'\, \frac{e^{E'}}{1+e^{\beta(E'-E)}}
=       e^E \int_0^{e^{-E}}\! \frac{dz}{1+z^\beta} 
\approx e^E \int_0^\infty\! \frac{dz}{1+z^\beta} = \cT\, e^E
\label{Gamma}
\ee
with $\cT = \pi T/\sin(\pi T)$. For Metropolis rates one has similarly
\be
\fl
\Gamma(E) = 
\int_{-\infty}^E\!dE'\,e^{E'}+\int_{E}^0\!dE'\,e^{E'}e^{-\beta(E'-E)}\approx
\int_{-\infty}^E\!dE'\,e^{E'}+\int_{E}^\infty\!dE'\,e^{E'}e^{-\beta(E'-E)}
\label{GammaM}
\ee
giving $\Gamma(E)=\cM e^E$ with
\be
\cM = 1+\frac{1}{\beta-1}=\frac{\beta}{\beta-1}=\frac{1}{1-T}
\ee
If the timescales in the system are
set by its age at long times, then typical values of $\Gamma(E)$
-- and therefore of $e^E$ -- at time $t$ should be of order $t^{-1}$.
This suggests the scaling ansatz $P(E,t) = e^E t \Psc(e^E t)$ where
$\Psc(\om)$ is the normalized probability distribution of $\om=e^E t$. The
zero-field master equation\eq{master_h0} for $P(E,m,t)$ gives after
integration over $m$
\be
\dd{t} P(E,t) = -\Gamma(E)P(E,t) + \int\! dE'\, \rho(E) w(E-E') P(E',t)
\ee
Inserting the scaling form for $P$ and neglecting the upper
cutoff $E'=0$ in the integral transforms this into an
integro-differential version for $\Psc(\om)$,
\be
\om \Psc'(\om) = -(1+\cT\,\om)\Psc(\om) + \int\!d\om'\,
\frac{\Psc(\om')}{1+(\om/\om')^\beta}
\label{Psc}
\ee
For Metropolis rates one only needs to replace $\cT$ by $\cM$ and
$[1+(\om/\om')^\beta]^{-1}$ by $\min\{(\om/\om')^{-\beta},1\}$, and
all statements for Glauber rates below can be translated to the Metropolis
case in an analogous way unless specified otherwise.

From\eq{Psc} it follows directly for $\om\to 0$ that $\Psc(0)=1$, as
also found by Bertin~\cite{Bertin03} who used $1/\om$ as the
scaling variable. For large $\om$, on the other hand, i.e.\
relatively shallow traps, one expects effective equilibration and
therefore $\Psc(\om)\sim e^{-\beta E} \sim \om^{-\beta}$. (There is no
density of states factor here because the density of states with
respect to $e^E$ is uniform on the allowed range $0<e^E<1$.) At $T=0$
this power-law 
tail becomes an exponential as we will see below. For a numerical
solution at $T>0$, it is useful to rewrite\eq{Psc} as
\be
\om \Psc(\om) = \int_0^\om\!d\om'\,f(\om')e^{-\cT(\om-\om')}, \qquad
f(\om) = \int\!d\om'\,\frac{\Psc(\om')}{1+(\om/\om')^\beta}
\label{Psc_numerical}
\ee
For later we also note the following relation between successive
moments of $\Psc(\om)$, obtained by multiplying\eq{Psc} with $\om^n$
($-1<n<\beta-1$) and integrating over $\om$:
\be
\frac{\int\!d\om\,\om^{n+1}\Psc(\om)}{\int\!d\om\,\om^n\Psc(\om)}
= n\left(\cT - \int\!dz\,\frac{z^n}{1+z^\beta}\right)^{-1}
= \frac{n}{\cT - \cTT{n}}
\label{P_moments}
\ee
The constant $\cTT{n}$ in this expression generalizes
$\cT\equiv\cTT{0}$:
\be
\cTT{n} = \int\!dz\,\frac{z^n}{1+z^\beta} = \frac{\pi T}{\sin[\pi T(n+1)]}
\ee
For Metropolis rates the analogous expression for the moment
ratio\eq{P_moments} is $n/(\cM-\cMM{n})$ with
\be
\cMM{n} = \int\! dz\,z^n\min\{z^{-\beta},1\} = \frac{\beta}{(n+1)(\beta-n-1)}
\label{cMn}
\ee

We next turn to the long-time behaviour of $\mb(E,t-\tw|\Ew)$. By
arguments similar to those above, this should have the scaling form
\be
\mb(E,t-\tw|\Ew) = \rh \mbsc(\rh,\ta), \qquad \rh=e^E/e^{\Ew}, \qquad
\ta = (t-\tw)e^{\Ew}
\ee
where $\mbsc(\rh,\ta)$ is the normalized distribution of $\rh$ after
the rescaled time-interval $\ta$.
%
%
The scaling form for $M$ then follows directly as
\be
\fl M(t-\tw|\Ew) = \int\!dE\,\mb(E,t-\tw|\Ew) = \Msc(\ta), \qquad
\Msc(\ta) = \int\!d\rh\,\mbsc(\rh,\ta)
\label{Msc}
\ee
The dependence on $\ta=(t-\tw)e^{\Ew}$ makes sense because, for
$|\Ew|\gg 1$, $e^{\Ew}$ sets the scale of the exit rate from traps of
depth $\Ew$. The dynamical equation\eq{M_eq} becomes in the scaling
regime
\be
\Msc'(\ta) = -\cT \Msc(\ta)
+ \k \int\! \frac{d\ta'}{\ta}\, \frac{\Msc(\ta')}{1+(\ta'/\ta)^\beta}
\label{Msc_eq}
\ee
This is similar in form to\eq{Psc} for $\Psc(\om)$; accordingly, there
is again an alternative version suitable for numerical iteration,
\be
\Msc(\ta) = e^{-\cT\ta} +
\k\int_0^\ta\!d\ta'\,g(\ta')e^{-\cT(\ta-\ta')}, 
\qquad
g(\ta) = \int\!\frac{d\ta'}{\ta}\,\frac{\Msc(\ta')}{1+(\ta'/\ta)^\beta}
\label{Msc_numerical}
\ee

The scaling of the correlation and response functions can now be
deduced. For the correlation\eq{corr} one gets
\be
C(t,\tw) = \Csc((t-\tw)/\tw), \qquad \Csc(x) =
\int\!d\om\,\Msc(x\om)\Psc(\om)
\label{corr_sc}
\ee
which displays the expected simple aging scaling with $x=(t-\tw)/\tw$,
implying that relaxation time scales grow linearly with the
age. Similarly, the scaling form of the response function\eq{resp} is
$R(t,\tw)=\tw^{-1}\Rsc(x)$ with
\bea
T\Rsc(x) &=& (1-\k) \int\! d\om\,\Msc(x\om) r(\om)
\label{resp_sc}
\\
r(\om) &=& \int\! d\om'\,
\frac{\Psc(\om)+\Psc(\om')}{[1+(\om/\om')^\beta][1+(\om'/\om)^\beta]}
\label{r_initial_def}
\\
&=& T\cT\, \om\Psc(\om) + \int\! d\om'\,
\frac{\Psc(\om')}{[1+(\om/\om')^\beta][1+(\om'/\om)^\beta]}
\label{r_def}
\eea
where we have used the integral
\be
\fl\int\! \frac{d\om'}{[1+(\om/\om')^\beta][1+(\om'/\om)^\beta]} = 
\om \int\! \frac{dz\, z^\beta}{(1+z^\beta)^2} = 
T\om \int\! dz\, z\left(-\frac{d}{dz}\right)\frac{1}{1+z^\beta} = 
T\cT\,\om
\label{d}
\ee
For Metropolis rates, one has similarly
\bea
r(\om) &=& \int\! d\om'\,
\left[\Psc(\om) \Theta(\om'-\om)(\om'/\om)^{-\beta}
+\Psc(\om')\Theta(\om-\om')(\om/\om')^{-\beta}\right]
\label{rM_initial_def}
\\
&=& T\cM \om\Psc(\om) + \int^\om\! d\om'\,
\Psc(\om')(\om/\om')^{-\beta}
\label{rM_def}
\eea
We will need the asymptotic behaviour of $r(\om)$ below. For small
$\om$, one can approximate $\Psc(\om)$ in\eq{r_initial_def} by 1. But
the denominator ensures that $\om$ and $\om'$ are of the same order,
so the same argument can be applied to $\Psc(\om')$. This gives
\be
r(\om)=2T\cT\,\om
\label{r_small_om}
\ee
in the limit of small $\om$; for Metropolis rates one finds
$r(\om)=2T\cM\om$ instead. For large $\om$, on the other hand, where
$\Psc(\om)\sim \om^{-\beta}$, one has
\be
\fl r(\om) = \Psc(\om)\int\! d\om'\,
\frac{1+(\om/\om')^\beta}{[1+(\om/\om')^\beta][1+(\om'/\om)^\beta]}
= \Psc(\om)\int\! d\om'\, \frac{1}{1+(\om'/\om)^\beta} = \cT\,\om
\Psc(\om)
\label{r_simp}
\ee
giving the scaling $r(\om)\sim \om^{1-\beta}$; the Metropolis case
again differs only by $\cT\to \cM$.

The susceptibility or step response $\chi(t,\tw) = \int_{\tw}^t\!dt'\,
R(t,t')$ follows from\eq{resp_sc} as
\be
\chi(t,\tw) = \int_{\tw}^t\!dt'\, \frac{1}{t'}\,\Rsc((t-t')/t') =
\chi(x), \qquad \chi(x)=\int_0^{x} \frac{dx'}{1+x'}\Rsc(x')
\label{chi_sc}
\ee
and again depends only on $x=(t-\tw)/\tw$, exhibiting simple aging
scaling. Finally, the fluctuation-dissipation ratio (FDR) scales in
the same manner,
\be
X(t,\tw) = \frac{TR(t,\tw)}{(\partial/\partial\tw)C(t,\tw)}
 = \Xsc(x), \qquad
\Xsc(x)=-\frac{T\Rsc(x)}{(1+x){\Csc}^\prime(x)}
\label{Xsc}
\ee
We recall that $X=1$ corresponds to equilibrium, where the
fluctuation-dissipation theorem holds; out of equilibrium, it then
makes sense to define $\Teff=T/X$ as an effective
temperature~\cite{CugKur93,CugKur94,CugKurPel97,CriRit02b}. This
quantity allows a
straightforward physical interpretation only when it is
time-independent, at least within a given time-sector; the one of
interest here is $t-\tw\sim \tw$. Since the definition\eq{Xsc} can be
written as 
$X=-(\partial T\chi/\partial\tw)/(\partial C/\partial \tw)$, such
time-independence 
corresponds to a straight-line FD plot of $\chi(t,\tw)$ versus
$C(t,\tw)$. More generally, if $\tw$ is used as the parameter varying along
the curve~\cite{FieSol02,SolFieMay02,MayBerGarSol03} then $X/T$ is the
negative slope of such an FD plot.

\section{Standard \BM\ model at $T>0$}
\label{sec:kappa0}

We first consider the case $\k=0$, i.e.\ the standard \BM\ model, at
arbitrary temperature $T$ ($<\Tg=1$) and with an exponential density of
states. One then has, from\eq{kappa0},\eq{Gamma} and\eq{Msc},
$\Msc(\ta)=\exp(-\cT\ta)$. The scaling forms\eq{corr_sc}
and\eq{resp_sc} of correlation and response function thus simplify to
\be
\Csc(x) = \int\!d\om\,e^{-\cT x\om}\Psc(\om)
\label{C_k0}
\ee
and
\be
T\Rsc(x) = \int\! d\om\, e^{-\cT x\om}r(\om)
\label{R_k0}
\ee
From this the large-$x$ asymptotics can be obtained directly: only the
region $\om\sim 1/x$ contributes to the integrals so that in $\Csc(x)$
we can set $\Psc(\om)\approx \Psc(0)=1$, giving
\be
\Csc(x) = \frac{1}{\cT x}
\ee
as found previously by Bertin~\cite{Bertin03}. For $\Rsc(x)$,
using\eq{r_small_om} leads similarly to
\be
T\Rsc(x) = \frac{2T}{\cT x^2}
\ee
The FDR from\eq{Xsc} thus behaves asymptotically as $\Xsc(x) =
2T/x$. In terms of $C$ this gives $X=-T\,d\chi/dC = 2T\cT\,C$ for small
$C$. The limiting FD plot therefore always starts parabolically at the
top, $\chi(C)=\chi(0)-\cT\,C^2$ for $C\to0$. The FDR itself decays
to zero asymptotically, $\Xsc(x\to\infty)=0$ for all $T<\Tg$, as in
the case of the Bouchaud trap model~\cite{BouDea95,FieSol02}.

More interesting is the short-time FDR $\Xsc(x\to 0)$. Naively, this
is from\eq{Xsc}
\be
\Xsc(x\to 0) = \frac{T\Rsc(0)}{-\Csc'(0)} = \frac{\int\! d\om\,r(\om)}
{\int\!d\om\,\cT\,\om \Psc(\om)}
=\frac{2T\cT\int\! d\om\,\om \Psc(\om)}{\cT\int\!d\om\,\om \Psc(\om)} = 2T
\label{X_smallT}
\ee
using\eq{C_k0} and\eq{R_k0} with $x=0$ as well as\eq{r_def} and\eq{d}
(or\eq{rM_def} for the Metropolis case). The effective temperature
associated with this FDR is therefore $\Teff=1/2$, independently of
$T$. However, this conclusion only holds for $T<1/2$. For larger $T$,
the integral
\be
-\Csc'(x) = \int\!d\om\,\cT\,\om e^{-\cT x\om}\Psc(\om)
\ee
is actually divergent for $x=0$ because of the
$\om^{-\beta}$ tail of $\Psc(\om)$. For small but nonzero $x$, the
exponential acts as a cutoff at $\om\sim 1/x$ so that
$-\Csc'(x)\sim(1/x)^{2-\beta}=x^{\beta-2}$, in agreement with the
short-time singularity in $C$ identified in~\cite{Bertin03}. The
response $T\Rsc(x) = \int\! d\om\, e^{-\cT x\om}r(\om)$ has the same
small-$x$ singularity because $r(\om)\sim \om^{1-\beta}$ for
$\om\to\infty$. In fact, for small $x$ one can replace $r(\om)$ by its
asymptotic form\eq{r_simp} to get
\be
T\Rsc(x) = \int \! d\om\, \cT\,\om e^{-\cT x\om}P(\om) = -\Csc'(x)
\ee
This shows that the short-time FDR is, for $T>1/2$,
\be
\Xsc(x\to 0) = 1
\label{X_largeT}
\ee
In this temperature-range the FD relation between $\chi$ and $C$ is
therefore always of a pseudo-equilibrium form in its initial part,
with a (negative) slope equal to $1/T$. This matches continuously with
the constant slope of $1/\Teff=2$ found above for $T<1/2$.

Intuitively, one can understand the occurrence of pseudo-equilibrium
behaviour for $T>1/2$ by looking at the average hopping rate. This is
\be
\Gamma(t) = \int\!dE\,\Gamma(E)P(E,t) = t^{-1}\int\!d\om \, \cT\,\om \Psc(\om)
\label{Gamma_asympt}
\ee
where the second form applies in the scaling regime. For $T<1/2$, the
integral converges and $\Gamma(t)\sim 1/t$ as one would naively expect
from our scaling assumption: typical relaxation times are $\sim t$,
thus typical rates are $\sim 1/t$. For $T>1/2$, however, the integral
is divergent at the upper end and one has to take into account the
cutoff at $E=0$, corresponding to $\om=t$, leading to $\Gamma(t)\sim
t^{1-\beta}$. This is entirely dominated by the very small probability
of being in shallow traps with atypically fast relaxation rates of
$\order(1)$. Now the response $R(t,\tw)$ and the initial decay
$(\partial C/\partial\tw)(t,\tw)$ of the correlation function are
sensitive only to hops taking place between $\tw$ and
$t$. For small $t-\tw$ these are precisely the same events that
dominate $\Gamma(t)$. Since they are in the ``effective equilibrium''
tail of $\Psc(\om)$, it is then not surprising to find a
pseudo-equilibrium form of the FD relation.

\begin{figure}
\begin{center}
\epsfig{file=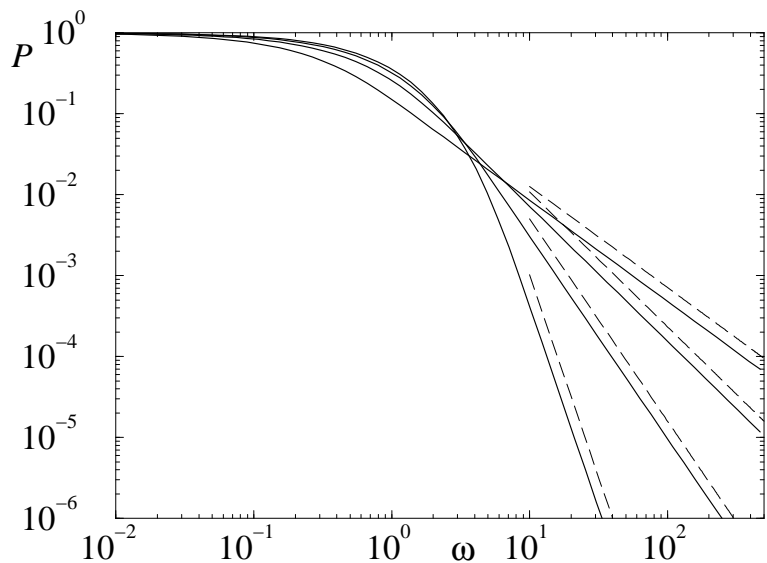,width=10cm}
\end{center}
\caption{Numerically calculated scaling distributions $\Psc(\om)$ for
Glauber dynamics at $T=0.2$, 0.4, 0.6, 0.8 (bottom to top on the
right). The dashed lines indicate the expected asymptotic scaling
$\Psc(\om)\sim\om^{-\beta}$.
\label{fig:P}
}
\end{figure}
Fig.~\ref{fig:P} shows some numerically calculated scaling
distributions $\Psc(\om)$ for Glauber rates; these exhibit the
expected asymptotic behaviour $\sim\om^{-\beta}$ for large $\om$. To
avoid end effects in the iterative numerical solution
of\eq{Psc_numerical}, we stored not $\Psc$ and $f$ themselves but
$\Psc(\om)(1+\om)^{\beta}$ and similarly
$f(\om)(1+\om)^{\beta-1}$. These functions have nonzero limits for
$\om\to\infty$ (and $\om\to 0$) and so are suitable for evaluating the
required integrals over $\om=0\ldots\infty$.

\begin{figure}
\begin{center}
\epsfig{file=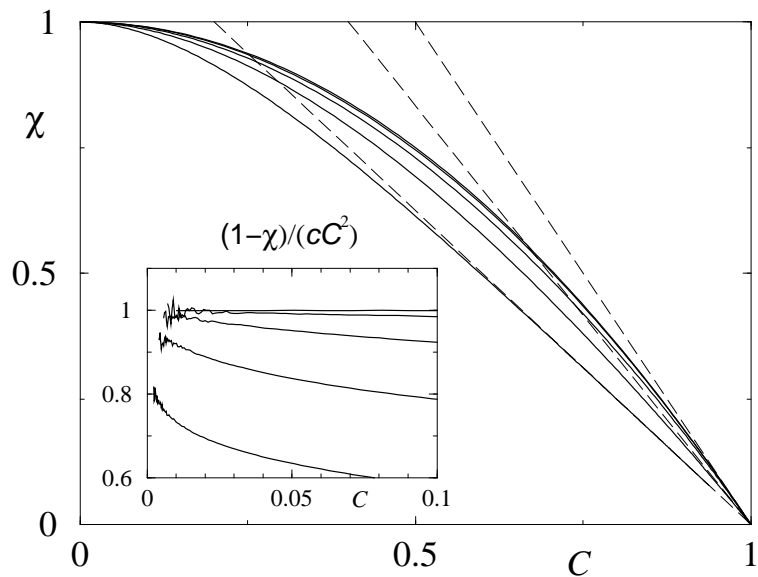,width=10cm}
\end{center}
\caption{Numerically calculated long-time FD relations $\chi(C)$ in
the original \BM\ model ($\k=0$, Glauber dynamics). Solid lines are
for $T=0$ (exact~\cite{Sollich03}), 0.2, 0.4, 0.6 and 0.8 from top to
bottom; the results for $T=0$ and $T=0.2$ are almost
indistinguishable. The dashed lines show the theoretically predicted
initial slopes (2 for $T<1/2$, and $1/T$ for $T>1/2$). The inset
graphs $(1-\chi)/(\cT\,C^2)$ versus $C$; our theory predicts that this quantity
converges to 1 for $C\to 0$.
\label{fig:FD_k0}
}
\end{figure}
Fig.~\ref{fig:FD_k0} displays the resulting FD plots of $\chi$ vs $C$;
these are valid in the limit of long times which we have already taken
by working in the scaling regime.
(To get reliable results for $\chi(x)$, one has to do
the $x'$-integration from\eq{chi_sc} {\em before} the
$\om$-integration in\eq{resp_sc}; otherwise the singularity of
$\Rsc(x)$ for $x\to 0$ at $T>1/2$ leads to problems.) The initial
slopes agree well with the theoretical predictions, as shown by the
dashed lines. The asymptotic slopes for $C\to 0$ are likewise
consistent with the predicted value of 0. The inset of the figure
explores this region in more detail, showing $(1-\chi)/(\cT\,
C^2)$. From the analysis above this quantity should converge to 1 for
$C\to 0$. The numerical data are consistent with this, though for
$T=0.8$ the approach to the limit is very slow. This makes sense: as
$T$ tends to $\Tg=1$ from below, the FD plot approaches a straight
line, and so the quadratic expansion around $C=0$ will be valid in a
region that shrinks to zero in the limit.

One notable feature of the numerical results is that the asymptotic
value $\chi(C=0)$ is equal to 1 for all $T<1$, at least to within our
numerical accuracy of around $10^{-5}$. This is as in the Bouchaud
model~\cite{Ritort03}, and has been conjectured also for the \BM\
model, on the same physical grounds~\cite{Sollich03}: the
susceptibility freezes to its value at $T=\Tg=1$ as $T$ is decreased
below the glass transition. It ought to be possible to confirm this
result analytically, but we have not yet found a way of doing this.

It should be noted that the FD plot is not quite $T$-independent for
$T<1/2$. While such $T$-independence had been suggested by the simulations
of~\cite{JunBer04}, the scaling-regime numerics shown in
Fig.~\ref{fig:FD_k0} clearly rule it out. Also, after a little reflection
one sees that higher derivatives of $C(x)$ diverge for $x\to 0$
already at lower temperatures; e.g.\ $C''(x)$ diverges for $T>1/3$,
$C'''(x)$ for $T>1/4$ etc.

\section{Slowly decorrelating observables, $T=0$}
\label{sec:T0}

In this section and the next we consider slowly decorrelating observables,
i.e.\ $\k>0$. We begin by analysing the zero temperature dynamics,
where a number of results can be derived for a general density of
states $\rho(E)$ and finite times (i.e.\ without taking the scaling
limit).

Compared to the standard case $\k=0$, the additional task is to
calculate $\mb(E,t-\tw|\Ew)$. At $T=0$, one has $w(E-E')=\Theta(E'-E)$
and $\Gamma(E)=\int_{-\infty}^E dE'\,\rho(E')$ for both Glauber and
Metropolis rates; equation\eq{mb_LT} thus simplifies to
\be
\frac{s+\Gamma(E)}{\rho(E)}\hat{\mb}(E,s|\Ew) -
\frac{1}{\rho(\Ew)}\delta(E-\Ew) = 
\k \int_E^0\! dE'\, \hat{\mb}(E',s|\Ew)
\ee
Differentiating w.r.t.\ $E$ gives a differential equation for
$\mb(E,s|\Ew)$ as a function of $E$. This is easily solved, with the result
\bea
\hat{\mb}(E,s|\Ew) &=& - \frac{1}{s+\Gamma(\Ew)} \frac{\partial}{\partial E}
\left[\Theta(\Ew-E)\exp\left(\k\int_E^{\Ew}\!
dE'\,\frac{\rho(E')}{s+\Gamma(E')}\right)\right]
\nonumber\\
&=& - \frac{1}{s+\Gamma(\Ew)} \frac{\partial}{\partial E}
\left[\Theta(\Ew-E)\left(\frac{s+\Gamma(\Ew)}{s+\Gamma(E)}\right)^\k
\right]
\label{mb_T0}
\eea
Before we exploit this to obtain $M$ and thence the correlation and
response functions, it is worth noting from\eq{mb_forward} that, for
the limiting case 
$\k=1$, $\mb(E,t-\tw|\Ew)$ is simply the propagator $P(E,t-\tw|\Ew)$
of the original \BM\ model without a field. Inverting the LT
in\eq{mb_T0} thus yields the exact zero-temperature propagator
\be
P(E,t-\tw|\Ew) = - \frac{\partial}{\partial E}
\left[\Theta(\Ew-E)e^{-\Gamma(E)(t-\tw)}\right]
\ee
Applying this to $\tw=0$ gives as the general solution starting from
an initial distribution $P(E,0)$
\be
P(E,t) = - \frac{\partial}{\partial E}
\left[e^{-\Gamma(E)t} \int_E^0\! dE' P(E',0)\right]
\ee
At long times most of the mass of this is a low $E$, where the
integral inside the square brackets can be set to one. One thus
recovers $P(E,t)=\rho(E)t\exp[-\Gamma(E)t]$ as the long-time scaling
form of the distribution over trap energies, independently of the
initial distribution~\cite{BarMez95}. For an exponential density of
states, $\rho(E)=e^E$, this becomes $P(E,t)=e^E t \Psc(e^E t)$ with the
scaling function $\Psc(\om)=\exp(-\om)$. This has an exponential tail
as anticipated above, and one easily checks that it solves the $T\to
0$ limit of\eq{Psc}.

Returning now to the result\eq{mb_T0} for general $\k$, we obtain for
the LT of $M(t-\tw|\Ew)$
\be
\fl \hat{M}(s|\Ew) = \int\!dE\, \hat{\mb}(E,s|\Ew) = \frac{1}{s+\Gamma(\Ew)} 
\left(\frac{s+\Gamma(\Ew)}{s}\right)^\k = s^{-\k}[s+\Gamma(\Ew)]^{\k-1}
\label{hatM}
\ee
Inverting the LT gives a hypergeometric function,
\be
M(t-\tw|\Ew) = {}_1F_1\biggl( 1-\k,1;-\Gamma(\Ew)(t-\tw) \biggr)
\label{M_T0}
\ee
If we now specialize to the exponential distribution of trap energies,
we have $\Gamma(E)=e^E$ and so\eq{M_T0} is exactly of the scaling
form\eq{Msc} discussed above, with
\be
\Msc(\ta) = {}_1F_1\biggl( 1-\k,1;-\ta\biggr)
= \sum_{k=0}^\infty\frac{(k-\k)!}{(-\k)!k!^2}(-\ta)^k
\label{Msc_T0}
\ee
(Non-integer factorials are defined in terms of the Gamma function,
$a! = \Gamma(a+1)$; we avoid the use of $\Gamma$ for such constants to
prevent confusion with our exit rates $\Gamma(E)$.) Both of these
results can also be obtained directly from the relevant
equations\eq{M_eq} and\eq{Msc_eq} for $M(t|\Ew)$ and $\Msc(\ta)$, by
representing these quantities as power series in $t$ and $\ta$,
respectively, and determining the coefficients recursively.

From $\Msc(\ta)$ we get the scaling limit of the correlation function,
using $\Psc(\om)=\exp(-\om)$,
\be
\Csc(x) = 
\int\! d\om\, \Msc(x \om) \Psc(\om) =
\sum_{k=0}^\infty\frac{(k-\k)!}{(-\k)!k!}(-x)^k = (1+x)^{\k-1}
\ee
so that $C(t,\tw)$ has the simple long-time form
$C(t,\tw)=(\tw/t)^{1-\k}$. This 
result can also be derived without recourse to series expansions:
denoting $\hat{\Msc}(\cdot)$ the LT of $\Msc(\ta)$, we have
from\eq{corr_sc}
\be
\int\!dx\,e^{-\sigma x}\Csc(x) = 
\int\!d\om\, \om^{-1}\hat{\Msc}(\sigma/\om)e^{-\om} =
\int\!dx\, x^{-1}\hat{\Msc}(x^{-1})e^{-\sigma x}
\ee
so that $\Csc(x)=x^{-1}\hat{\Msc}(x^{-1})$. With $\hat{\Msc}(x^{-1}) =
(x^{-1})^{-\k}(x^{-1}+1)^{\k-1}$ from\eq{hatM} one gets
$\Csc(x)=(1+x)^{\k-1}$ as before.

\begin{figure}
\begin{center}
\epsfig{file=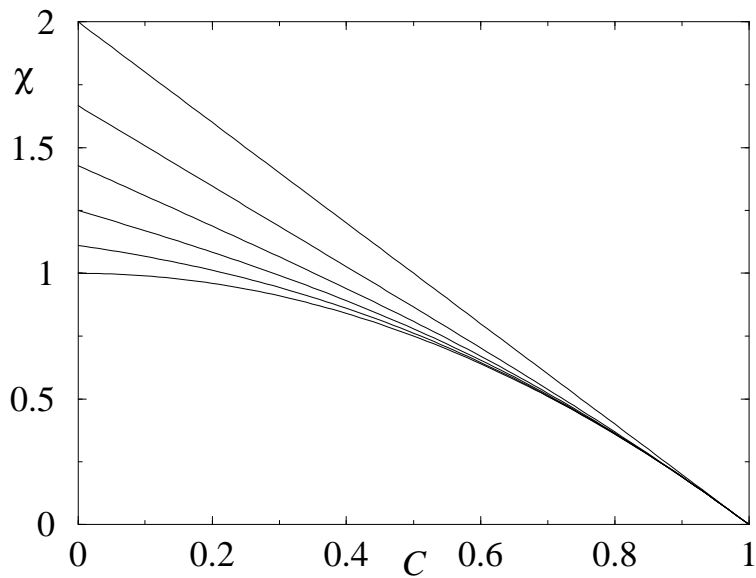,width=10cm}
\end{center}
\caption{Long-time FD plots $\chi(C)$ at $T=0$, for $\k=0$, 0.2, 0.4,
0.6, 0.8, 1 from bottom to top.
\label{fig:T0_FD}
}
\end{figure}
For the response, we note that the denominator factor
in\eq{r_initial_def} tends to $T\om\delta(\om-\om')$ for $T\to 0$; the
same conclusion also holds in the Metropolis
case\eq{rM_initial_def}. This gives for the scaling function
\bea
\fl \Rsc(x) &=& 2(1-\k) \int\! d\om\, \Msc(x\om)\om \Psc(\om)
= 2(1-\k) \sum_{k=0}^\infty\frac{(k+1)(k-\k)!}{(-\k)!k!}(-x)^k
\\
\fl &=& 2(1-\k)\frac{\partial}{\partial x}[x\Csc(x)] = 
2\k(1-\k)(1+x)^{\k-1} + 2(1-\k)^2(1+x)^{\k-2}
\eea
For the susceptibility this implies, from\eq{chi_sc},
\bea
\chi(x) &=& 2\k\left[1-(1+x)^{\k-1}\right]
+ \frac{2(1-\k)^2}{2-\k}\left[1-(1+x)^{\k-2}\right]
\\
&=& 2\k\left[1-C\right]
+ \frac{2(1-\k)^2}{2-\k}\left[1-C^{(2-\k)/(1-\k)}\right]
\eea
where in the second equality $x$ has been expressed in terms of $C$ to
give the long-time FD relation $\chi(C)$. This is displayed in
Fig.~\ref{fig:T0_FD} for some representative values of $\k$. The slope
of the FD plot is $-\chi'(C)=2\k+2(1-\k)C^{1/(1-\k)}$, with
initial value $-\chi'(C=1)=2$ independently of $\k$, corresponding to
an ``effective short-time temperature'' of $\Teff=1/2$ ($=\Tg/2$ in
dimensional units). The asymptotic slope, on the other hand, is
$-\chi'(C=0)=2\k$ and depends continuously on $\k$. In the limit
$\k\to 1$ (where $C(x)=(1+x)^{\k-1}$ decays very slowly with $x$, and
$\chi(x)$ grows correspondingly slowly), the FD plot becomes a
straight line of negative slope 2, suggesting that this slowly
decorrelating observable measures a well-defined $\Teff=1/2$. This
result lends support to the straight-line FD plots found in the
simulations of \JB~\cite{JunBer04}.

For later, we note that by expressing the ratio of factorials
in\eq{Msc_T0} as a Beta function integral and then performing the sum
over $k$ one gets an alternative form for $\Msc(\ta)$,
\be
\Msc(\ta) =
\frac{1}{(-\k)!(\k-1)!}\int_0^1\!dz\,e^{-z\ta}z^{-\k}(1-z)^{\k-1}
\ee
This implies in particular that $\Msc(\ta)=\ta^{-(1-\k)}/(\k-1)!$ for
large $\ta$ and $\k>0$.

\section{Slowly decorrelating observables, $T>0$}
\label{sec:kappa_T}

Finally we consider the most general case of dynamics at nonzero
temperature and observables with incomplete decorrelation ($\k>0$). We
focus directly on the scaling limit for a system with an exponential
density of states; the key issue is again solving for $\Msc(\ta)$.
For short times this is easy: from\eq{Msc_eq}, $\Msc'(0)=-\cT+\k\cT$
and thus $\Msc(\ta)=1-(1-\k)\cT\ta+\order(\ta^2)$. From this the
initial FDR for $T<1/2$ can be worked out as in\eq{X_smallT}
\be
\Xsc(x\to 0) = \frac{T\Rsc(0)}{-\Csc'(0)} = 
\frac{(1-\k)\int\! d\om\,r(\om)}{\int\!d\om\,(1-\k)\cT\,\om \Psc(\om)}
= 2T
\ee
The $\k$-dependent factors $(1-\k)$ cancel, so the initial slope of
the FD plot remains 2 for all $\k$. Metropolis rather than Glauber
rates again only replace $\cT$ by $\cM$ everywhere.

For $T>1/2$, the linearization in $\ta$ breaks down as before, and
the integrals for $\Rsc(x)$ and $1-\Csc(x)$ will be dominated by large
$\om\sim 1/x$ for $x\to 0$. One can therefore use the asymptotic
behaviour of $\Psc(\om)\propto \om^{-\beta}$ and $r(\om)\propto
\cT\,\om^{1-\beta}$ (with the same proportionality constant, according
to\eq{r_simp}) to write
\bea
\fl 1-\Csc(x) &\propto& \int\!d\om\,[1-\Msc(x\om)]\om^{-\beta} 
= x^{\beta-1}\int\!d\ta\,[1-\Msc(\ta)]\ta^{-\beta}
\\
\fl T\Rsc(x) &\propto& (1-\k)\cT\int\!d\om\,\Msc(x\om)\om^{1-\beta} 
= (1-\k)\cT x^{\beta-2}\int\!d\ta\,\Msc(\ta)\ta^{1-\beta}
\eea
These singularities (but not their prefactors) are the same as for
$\k=0$, and correspondingly the FDR again approaches a nonzero limit
for $x\to 0$:
\be
\fl \Xsc(x) = -\frac{T\Rsc(x)}{(1+x)\Csc'(x)} \to
\frac{(1-\k)\cT \int\!d\ta\,\Msc(\ta)\ta^{1-\beta}}
{(\beta-1) \int\!d\ta\,[1-\Msc(\ta)]\ta^{-\beta}}
= \frac{(1-\k)\cT \int\!d\ta\,\Msc(\ta)\ta^{1-\beta}}
{-\int\!d\ta\,\Msc'(\ta)\ta^{1-\beta}}
\label{X0_largeT}
\ee
At first sight the value of $\Xsc(x\to 0)$ appears to depend on the
precise functional form of $\Msc(\ta)$. But in fact, by
multiplying\eq{Msc_eq} by $\ta^{1-\beta}$ and integrating over $\ta$ one
deduces 
\be
\fl \int\!d\ta\,\Msc'(\ta)\ta^{1-\beta} = -\cT
\int\!d\ta\,\Msc(\ta)\ta^{1-\beta}
+ \k \int\! d\ta'\, \Msc(\ta')
\int\!\frac{d\ta}{\ta}\, \frac{\ta^{1-\beta}}{1+(\ta'/\ta)^\beta}
\ee
The last $\ta$-integral evaluates to $\cT(\ta')^{1-\beta}$ and this
implies from\eq{X0_largeT} that $\Xsc(x\to 0)=1$ for all $\k$ (and
$T>1/2$). Physically, this result---which can be derived similarly for
Metropolis dynamics---is supported by the same intuition as for $\k=0$:
the initial response and decay of the correlation, and hence the FDR,
are dominated by the very small fraction of histories which pass
around time $\tw$ through
the shallow traps near the top of the energy landscape
($E=\order(1)$), where an 
effective pseudo-equilibrium is established.

What about long time intervals, $x\gg 1$? From the definition of the
dynamics, it is clear that $C(t,\tw)$ is the average of $\k^j=e^{j\ln\k}$ over
the distribution of the number of hops $j$ between $\tw$ and $t$. For
$\k\approx1$, a large number $j\sim 1/(-\ln
\k)\approx 1/(1-\k)$ of hops is needed to get any significant
decorrelation. Across the corresponding long time intervals the number
of hops should average out, leading to the naive prediction
$C\approx\exp[(\ln\k)\int_{\tw}^t\!dt'\,\Gamma(t')]$.
Using\eq{Gamma_asympt}, this gives the estimate
\be
\Csc(x)=(1+x)^{-\dT(1-\k)}, \qquad \dT = \cT \int\!d\om \, \om \Psc(\om)
\label{Csc_naive}
\ee
Intuitively, one expects that similarly $\Msc(\ta)\sim
\ta^{-\dT(1-\k)}$ for large $\ta$. The constant $\dT$ can be evaluated
explicitly, by taking the $n\to 0$ limit of\eq{P_moments}, giving
\be
\dT = \cT\, \frac{\sin^2(\pi T)}{(\pi T)^2\cos(\pi T)} = \frac{\tan(\pi
T)}{\pi T}
\ee
The analogous expression for the Metropolis case is, using\eq{cMn},
\be
\dM = \cM \int\!d\om\,\om\Psc(\om) =
\frac{\beta}{\beta-1}\frac{(\beta-1)^2}{\beta(\beta-2)} =
\frac{\beta-1}{\beta-2} = \frac{1-T}{1-2T}
\label{dM}
\ee
and together with\eq{Csc_naive} confirms the estimate for the
correlation function scaling in eq.~(22) of~\cite{JunBer04}. Both
$\dT$ and $\dM$ diverge as $T\to 1/2$ from below, signalling a
breakdown of the above naive reasoning for higher temperatures; we
return to this point below. For $T\to 0$, $\dT$ and $\dM$ both
approach unity so that $\Msc(\ta)$ is predicted to decay as
$\ta^{-(1-\k)}$, exactly as we found by exact calculation for $T=0$.

To find the asymptotics of $\Msc(\ta)$ for general $T$ and $\k$, one
substitutes the ansatz $\Msc(\ta)\sim\ta^{-\lam}$ into\eq{Msc_eq}. The
l.h.s.\ is then subleading and the leading terms on the r.h.s.\ have
to cancel, giving the condition
\be
\cT=\k\cTT{-\lam}
\label{lambda_cond}
\ee
for $\lam$, or equivalently $\k\sin(\pi T) = \sin[\pi T(1-\lam)]$. For
$\cTT{-\lam}$ to be finite one needs $\lam<1$, and $\Msc(\ta)$
decreases with $\ta$ so $\lam>0$.
\begin{figure}
\begin{center}
\epsfig{file=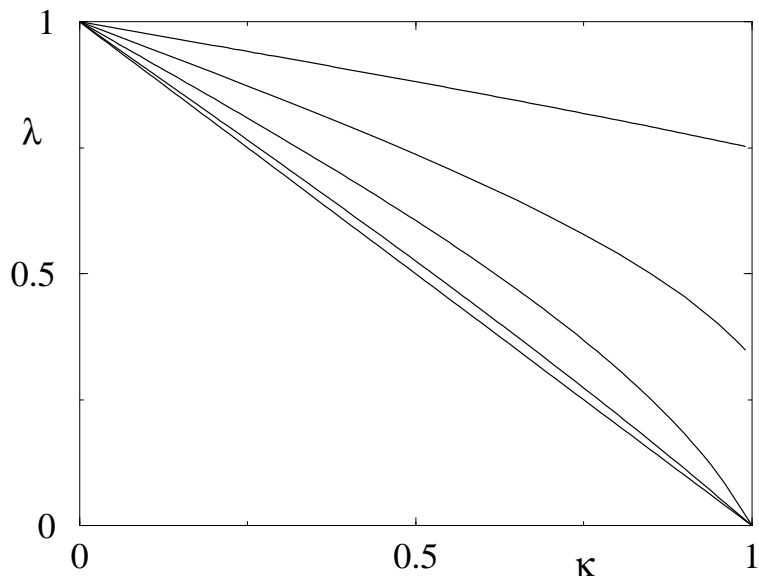,width=10cm}
\end{center}
\caption{Calculated values of the decay exponent $\lam$ of
$\Msc(\tau)\sim\tau^{-\lam}$ as a function of $\kappa$.  The curves
are for Glauber dynamics and $T=0$, 0.2, 0.4, 0.6, 0.8 from bottom to
top.
\label{fig:lambda_kappa}
}
\end{figure}
In this range of $\lambda$ the
condition\eq{lambda_cond} has only one solution,
\be
\lam=1-\frac{\arcsin[\k\sin(\pi T)]}{\pi T}
\label{lambda}
\ee
For small $\k$ this exponent approaches 1, with
$\lam=1-\k/\cT+\order(\k^2)$; see Fig.~\ref{fig:lambda_kappa}. The
behaviour for $\k\to 1$, on the other hand, depends strongly on
temperature. For $T<1/2$, $\lam=(1-\k)\tan(\pi T)/(\pi T)=\dT(1-\k)$
to first order in $1-\k$, exactly as the naive argument had
predicted. For higher temperatures $T>1/2$, the exponent approaches a
nonzero value $\lam=2-\beta$ in the same limit. This is rather
striking since {\em at} $\k=1$ directly one expects $\Msc(\ta)\equiv
1$ and hence $\lam=0$: the limit $\k\to 1$ is discontinuous for
$T>1/2$. We discuss the physical reasons for this at first sight
surprising behaviour below.

The above predictions for $\lam$ apply to the Glauber case. Using the
same arguments for Metropolis rates, one is led to the condition $\cM
=\k\cMM{-\lam}$. This reads explicitly $\beta/(\beta-1) =
\k\beta/[(1-\lam)(\beta+\lam-1)]$, giving
\be
\lam = 1-\frac{1}{2}\beta + \sqrt{\frac{1}{4}\beta^2 - \k(\beta-1)}
\label{lambdaM}
\ee
Similarly to the Glauber case, $\lam\to0$ for $\k\to 1$ at $T<1/2$,
with leading order behaviour $\lam=(1-\k)(\beta-1)/(\beta-2)$ in
agreement with\eq{dM}. For $T>1/2$, on the other hand,
$\lam\to2-\beta$ and this limit value is the same as for Glauber
dynamics. In general, however, as soon as the decorrelation of the
observable in a jump is not perfect ($\k>0$), the decay
exponents\eq{lambda} for Glauber and\eq{lambdaM} for Metropolis rates
are {\em different}. This provides an interesting example where
microscopic details of the assumed transition rates affect not only
quantitative details (prefactors) but also qualitative aspects
(exponents) of the dynamics.

The power-law decay of $\Msc(\ta)$ implies similar large-$x$
asymptotics for the
response and correlation functions, consistent with the naive
expectation explained above:
\bea
\Csc(x)&\propto& x^{-\lam} \int\!d\om\, \om^{-\lam} \Psc(\om)\\
T\Rsc(x)&\propto& (1-\k) x^{-\lam} \int\!d\om\, \om^{-\lam} r(\om)
\eea
In particular, for $T>1/2$ and $\k\to 1$, the correlation decays as
$\Csc(x) \sim x^{-\lam}$ with $\lam=2-\beta$. This exact value is
consistent with the argument by \JB~\cite{JunBer04} that the decay
exponent $\lam$ should be close to $T=1/\beta$ for $T\approx 1$; to first
order in $1-T$ this estimate agrees with the exact prediction. It deviates
progressively for lower $T$, however, with the exact value 
$\lambda=2-\beta$ dropping to zero at
$T=1/2$. This is as it must be by continuity, since for $T<1/2$ we
found that $\lambda$ vanishes for $\k\to 1$. As a check
of our theory we have also compared the predicted exponent\eq{lambdaM} for
Metropolis dynamics with the numerical correlation function results from
Fig.~9 of~\cite{JunBer04} for $\k=0.9$, and found very good agreement.

From the above large-$x$ behaviour of $\Csc$ and $\Rsc$ we can deduce
the asymptotic FDR\eq{Xsc} as
\be
\Xsc(x\to \infty) = \frac{(1-\k) \int\!d\om\, \om^{-\lam} r(\om)}
{\lam\int\!d\om\, \om^{-\lam} \Psc(\om)}
\ee
Using\eq{r_def}, one finds that the numerator integral equals
$T[\cT + (1-\lam)\cTT{-\lam}]\int\!d\om\,
\om^{1-\lam}\Psc(\om)$. The remaining moment ratio 
can then be calculated from\eq{P_moments} with $n=-\lam$ to give
\be
\Xsc(x\to \infty)
= \frac{(1-\k) T[\cT+(1-\lam)\cTT{-\lam}]} {\cTT{-\lam} - \cT}
\ee
Inserting the condition\eq{lambda_cond} determining $\lam$ simplifies this to
\be
\Xsc(x\to \infty) = 
T(1-\lam+\k)
\ee
One can check that this expression is valid also for Metropolis rates,
with the value of $\lam$ then given by\eq{lambdaM} instead of
$\lam$. For both Glauber and Metropolis rates the asymptotic FDR
$\Xsc(x\to\infty)$ 
approaches zero for $\k\to 0$, consistent with the results for the
ordinary \BM\ model from Sec.~\ref{sec:kappa0}. The behaviour for
$\k\to 1$ again depends on 
$T$. For $T<1/2$, $\lam\to 0$ and thus $X(x\to\infty) \to 2T$. For
$T>1/2$, on the other hand, $\lam\to 2-\beta$ and $X(x\to\infty)\to
1$. Both of these values are identical to the ($\k$-independent) {\em
short-time} FDR $\Xsc(x\to 0)$ found above. This strongly suggests that
the FD plots are straight lines for $\k\to 1$, both above and below
$T=1/2$. We now proceed to show this, by considering the FDR $\Xsc(x)$
for general $x$.

To this end, it will be useful to have an expression for $\Csc'(x)$
that is similar in form to $\Rsc(x)$. Starting from\eq{corr_sc} one
has $\Csc'(x) = \int\!d\om\,\Msc'(x\om)\om\Psc(\om)$. With\eq{Msc_eq}
this can be written in terms of $\Msc(\cdot)$ itself as
\be
\Csc'(x) = -\cT\int\!d\om\,\Msc(x\om)\om\Psc(\om) + 
\k\int\!d\om\, \Msc(x\om) \int\!d\om'\,
\frac{\Psc(\om')}{1+(\om/\om')^\beta}
\label{Csc_prime_aux}
\ee
The $\om'$-integral is, from\eq{Psc}, $(d/d\om)[\om\Psc(\om)] +
\cT\,\om\Psc(\om)$. The second term in this sum gives a contribution of
the same form as the first term on the l.h.s.\ of\eq{Csc_prime_aux},
and after an integration by parts
\be
\Csc'(x) = -\cT(1-\k)\int\!d\om\,\Msc(x\om)\om\Psc(\om) 
- \k x\int\!d\om\, \Msc'(x\om) \om \Psc(\om)
\ee
The last term is now just $-\k x\Csc'(x)$ and we end up with the
relatively simple expression
\be
-(1+\k x)\Csc'(x) = \cT(1-\k)\int\!d\om\,\Msc(x\om)\om\Psc(\om) 
\label{Csc_prime}
\ee
which also holds for Metropolis dynamics ($\cT\to \cM$). The
FDR\eq{Xsc} then becomes a ratio of integrals involving only
$\Msc(\cdot)$, rather than $\Msc(\cdot)$ and $\Msc'(\cdot)$ as before,
\be
\Xsc(x) = \frac{1+\k x}{1+x}
\frac{\int\!d\om\,\Msc(x\om)r(\om)}
{\cT\int\!d\om\,\Msc(x\om)\om\Psc(\om)}
\label{Xsc_simp0}
\ee
We can now see how this simplifies for $\k\to 1$. For $T<1/2$,
$\Msc(\ta)\to 1$ as $\k\to 1$ for any fixed $\ta$. The $\om$-integrals
remain finite in this limit, so that
\be
\Xsc(x) \to \frac{\int\!d\om\,r(\om)}{\cT\int\!d\om\,\om\Psc(\om)} = 2T
\label{Xsc_simp}
\ee
exactly as in\eq{X_smallT}: $\Xsc(x)=2T$ becomes independent of $x$,
and the FD plot is therefore a straight line of slope $X/T=2$. For
$T>1/2$, on the other hand, $\Msc(\ta)$ is asymptotically a power law
$\ta^{-\lam}$, with $\lam$ approaching $2-\beta$ from above as $\k\to
1$; thus for any fixed $x>0$ also $\Msc(x\om)\sim\om^{-\lam}$ for
large $\om$. But $r(\om)\sim \om\Psc(\om)\sim \om^{1-\beta}$ and so
the integrals in\eq{Xsc_simp0} become divergent at the upper end as
$\k\to 1$. They are therefore dominated by large $\om$-values, where
$r(\om) = \cT\,\om \Psc(\om)$. This shows that $\Xsc(x)\to 1$: the FD
plot is again a straight line, but now of equilibrium slope $1/T$.

With similar arguments we can also work out the correlation functions
for $\k\to 1$. Consider $T<1/2$ first, and divide\eq{Csc_prime} by
$\Csc(x)$ to get
\be
-(1+\k x)\frac{\Csc'(x)}{\Csc(x)} 
= \frac{\cT(1-\k)\int\!d\om\,\Msc(x\om)\om\Psc(\om)}
{\int\!d\om\,\Msc(x\om)\Psc(\om)}
\ee
In the limit $\k\to 1$ we can set $\Msc\to 1$ again and get
$-d(\ln\Csc)/d(\ln(1+x))=\cT(1-\k)\int\!d\om\,\om\Psc(\om)=\dT(1-\k)$.
This implies $\Csc(x)=(1+x)^{-\dT(1-\k)}$, exactly as the naive
argument\eq{Csc_naive} suggested

For $T>1/2$, consider the integral on the r.h.s.\
of\eq{Csc_prime}. This becomes divergent at the upper end for $\k\to
1$ as explained above, and dominated by large $\om$. $\Msc(x\om)$ can
therefore be replaced by its asymptotic form $\sim(x\om)^{-\lam}$, and
the r.h.s.\ of\eq{Csc_prime} becomes proportional to $x^{-\lam}$,
which in the limit $\k\to 1$ is $x^{\beta-2}$. (The remaining integral diverges
as $\k\to 1$, but combines with the $1-\k$ prefactor to give a finite
limit.) Thus, for $T>1/2$ and $\k\to 1$,
\be
-\Csc'(x) \propto \frac{x^{\beta-2}}{1+x}\ ,
\qquad
\Csc(x) = \frac{\sin[\pi(\beta-1)]}{\pi}
\int_x^\infty\!dx'\, \frac{(x')^{\beta-2}}{1+x'}
\label{k1_largeT}
\ee
In the expression for $\Csc(x)$ we have explicitly put in the
proportionality constant, which follows from the equal-time value
$\Csc(0)=1$. The same result also holds for Metropolis
dynamics, since both the limit value of $\lam$ for $\k\to 1$ and the asymptotic
behaviour of $\Psc(\om)$ and $r(\om)$, on which the argument relies,
are the same.

\subsection{Numerical results}

\begin{figure}
\begin{center}
\epsfig{file=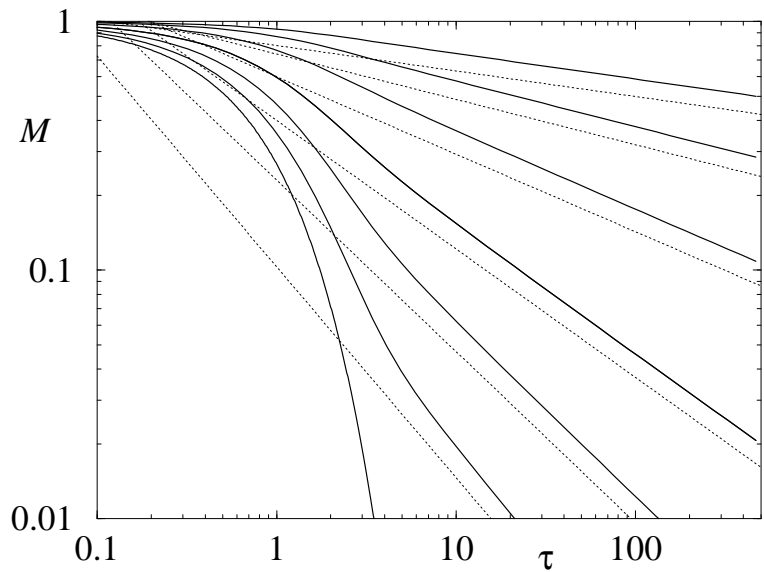,width=10cm}
\end{center}
\caption{Calculated $\Msc(\ta)$ for $T=0.4$ and $\k=0$ (exact), 0.2,
0.4, 0.6, 0.8, 0.9, 0.95, from bottom to top. Dotted lines indicate
the predicted $\ta^{-\lam}$ asymptotes.
\label{fig:T0.4_M}
}
\end{figure}
\begin{figure}
\begin{center}
\epsfig{file=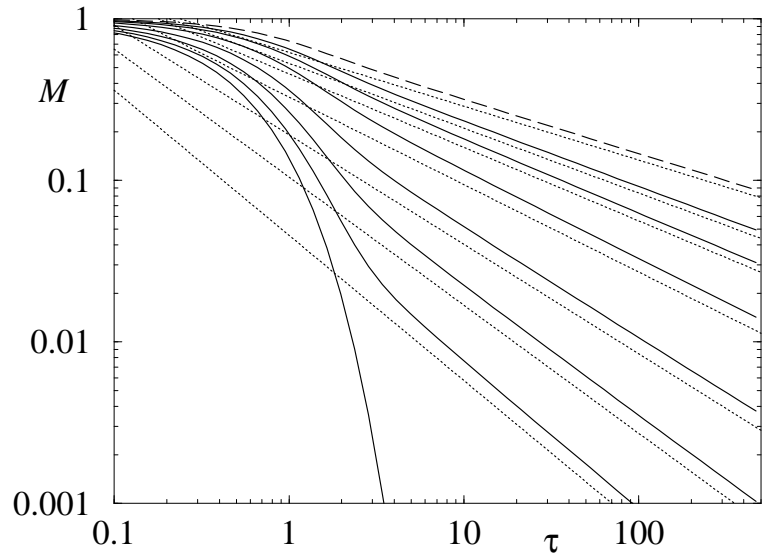,width=10cm}
\end{center}
\caption{Analogue of Fig.~\protect\ref{fig:T0.4_M} for $T=0.6$, for
the same values of $\k$. The dashed line shows in addition the
numerically calculated limiting form of $\Msc(\ta)$ for $\k\to 1$.
\label{fig:T0.6_M}
}
\end{figure}

We solved\eq{Msc_numerical} numerically to get $\Msc(\ta)$, using a
representation which takes into account the asymptotic behaviour
$\ta^{-\lam}$, and then evaluated correlation and response. The results for
$\Msc(\ta)$ at $T=0.4$ are shown in Fig.~\ref{fig:T0.4_M}, and do
approach the predicted asymptotes. In particular, the slope $\lam$ of
the asymptotic power law decreases to zero for $\k\to 1$, while for
$T=0.6$ (Fig.~\ref{fig:T0.6_M}) it approaches the nonzero limit
$2-\beta$. The dashed line shows the limiting form of $\Msc(\ta)$ for
$\k\to 1$, which for $T>1/2$ is nontrivial. Note that we focus
on Glauber dynamics throughout this section; graphs for the Metropolis
case would look broadly similar as discussed above.

\begin{figure}
\begin{center}
\epsfig{file=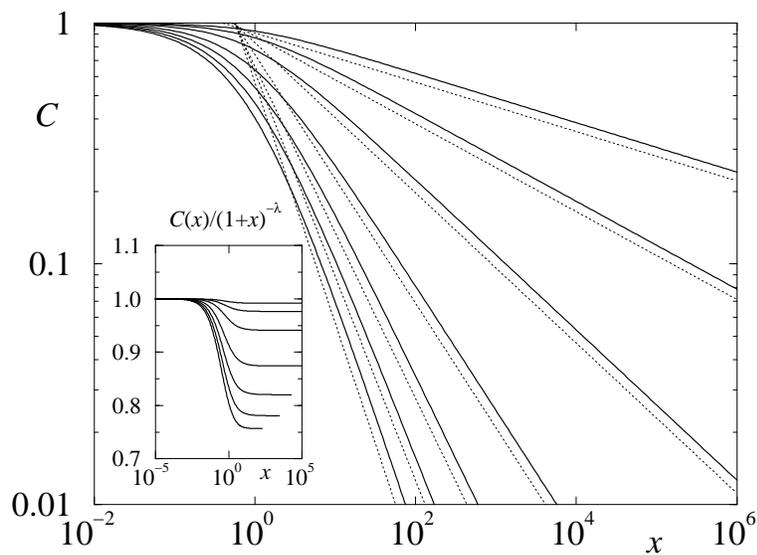,width=10cm}
\end{center}
\caption{Correlation function $\Csc(x)$ for $T=0.4$ and the same $\k$
as in Fig.~\protect\ref{fig:T0.4_M}. Dotted lines indicate the
predicted asymptotes $x^{-\lam}$. The inset shows
$\Csc(x)/(1+x)^{-\lam}$ versus $x$.
\label{fig:T0.4_C}
}
\end{figure}
\begin{figure}
\begin{center}
\epsfig{file=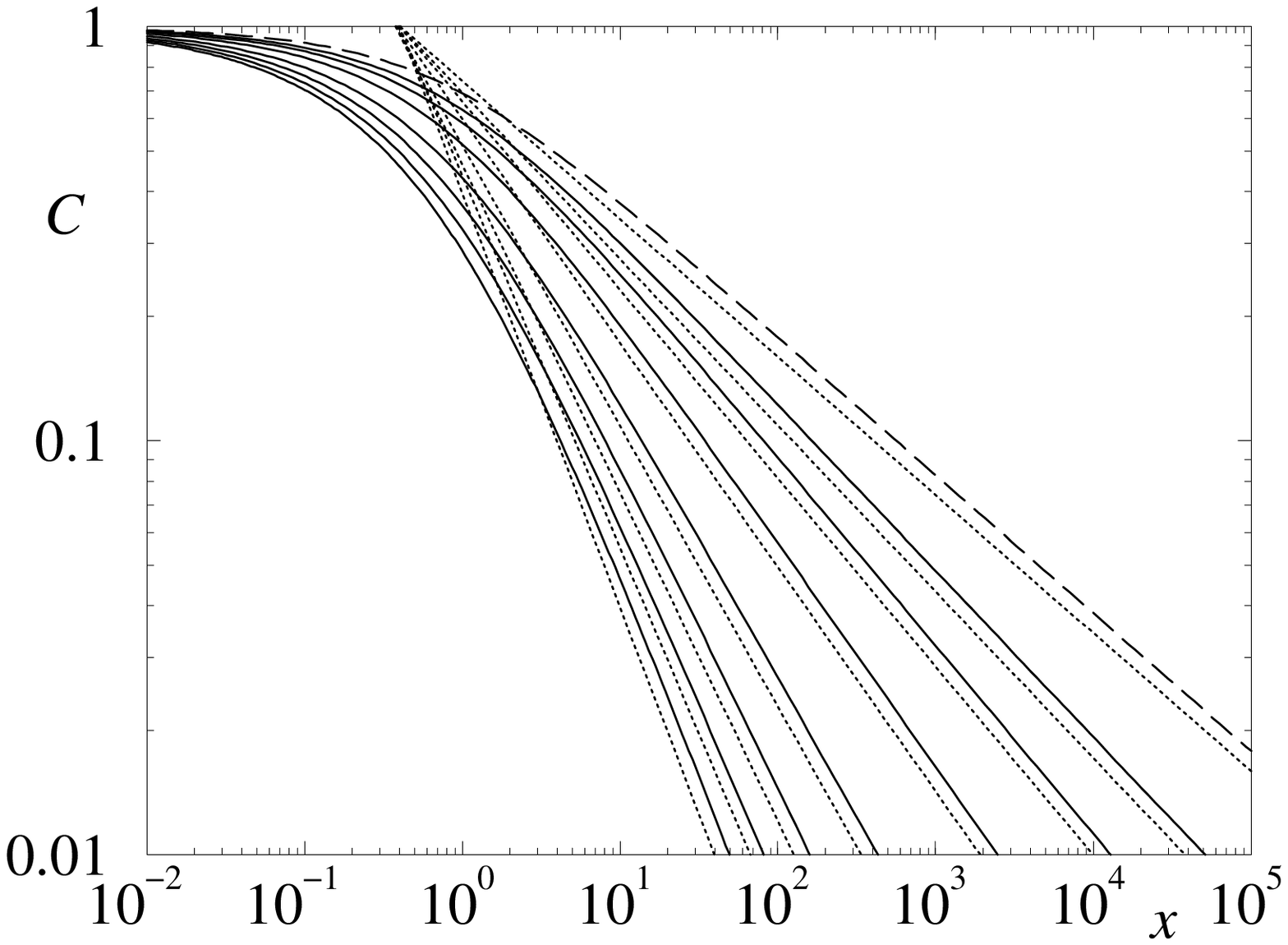,width=10cm}
\end{center}
\caption{Correlation function $\Csc(x)$ for $T=0.6$ and the same $\k$
as in Fig.~\protect\ref{fig:T0.6_M}; the dashed line shows the
predicted $\k\to 1$ limit\eq{k1_largeT}. Dotted lines indicate the
predicted asymptotes $x^{-\lam}$.
\label{fig:T0.6_C}
}
\end{figure}
Fig.~\ref{fig:T0.4_C} shows the scaling function $\Csc(x)$ of the
two-time correlation at $T=0.4$, again for a range of values of
$\k$. As expected, the asymptotic decay follows the same power law as
for $\Msc(\ta)$, i.e.\ $\Csc(x)\sim x^{-\lam}$. In the inset we show
$\Csc(x)/(1+x)^{-\lam}$ to demonstrate that this approaches unity in
the limit $\k\to 1$. Fig.~\ref{fig:T0.6_C} displays the analogous data
for $T=0.6$. The predicted asymptotic behaviour is again observed, but
now $\Csc(x)$ approaches the nontrivial limiting form for $\k\to 1$
predicted by\eq{k1_largeT} (dashed line).

\begin{figure}
\begin{center}
\epsfig{file=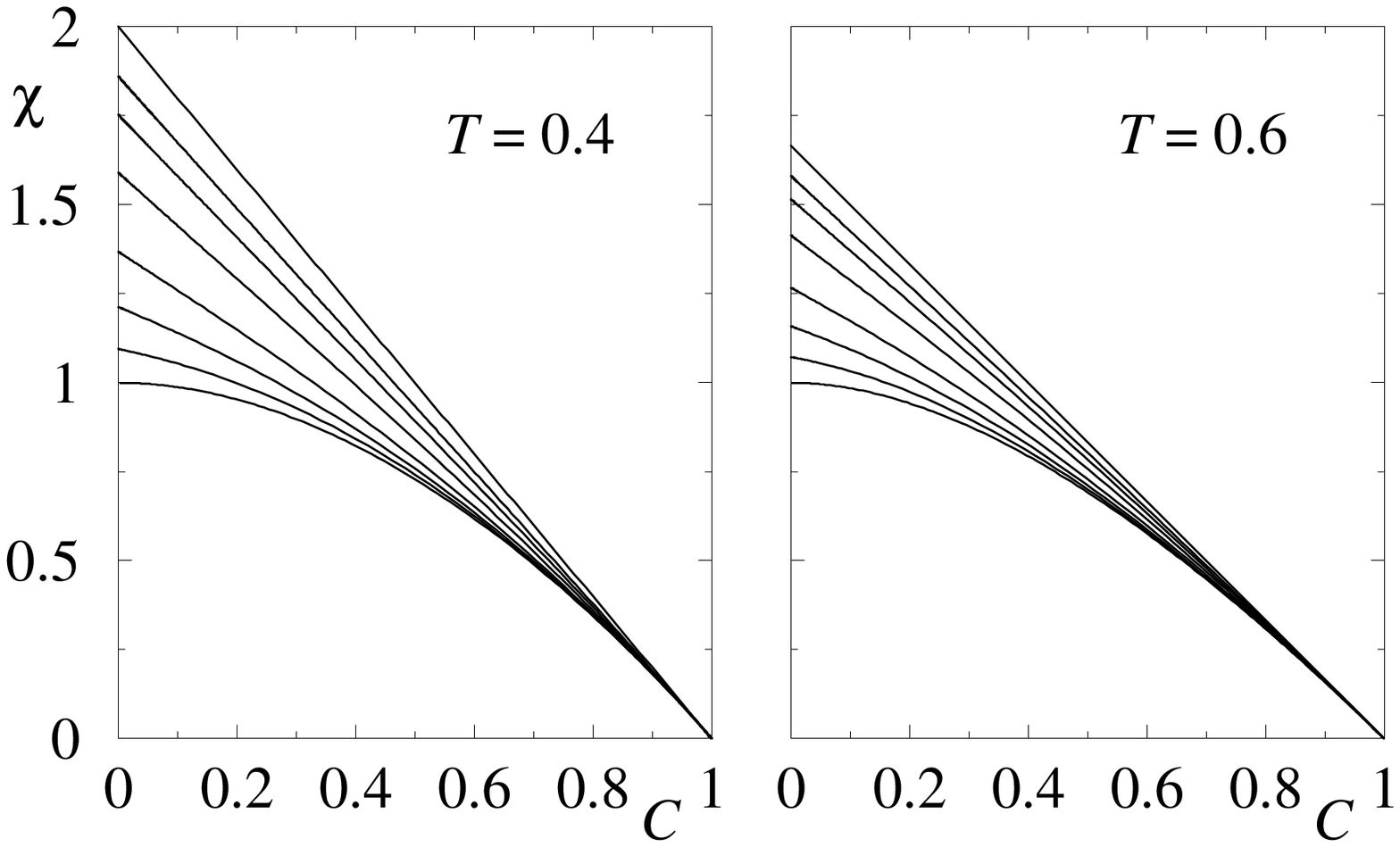,width=10cm}
\end{center}
\caption{FD plots for $\k=0$, 0.2, 0.4, 0.6, 0.8, 0.9, 0.95
1.0 (the last one being exact) from bottom to top, at temperature
$T=0.4$ (left) and $T=0.6$ (right).
\label{fig:T0.4_FD}
}
\end{figure}
%
%
%
The resulting FD plots (Fig.~\ref{fig:T0.4_FD}, left) for $T=0.4$ move
upwards with increasing $\k$ as for $T=0$ and are consistent with the
approach to the predicted straight line of slope $2$. For $T=0.6$
(Fig.~\ref{fig:T0.4_FD}, right) the overall trend is similar, but the
limiting straight line now has slope $1/T$, again as predicted.

\subsection{Discussion}
\label{sec:disc}

The results found above for the limit $\k\to 1$ of slowly
decorrelating observables provide strong support for the arguments of
\JB~\cite{JunBer04}, in particular in terms of the straight-line FD
plots and the asympotic decay of $\Csc(x)$ for $T<1/2$. In addition
they clarify the limiting behaviour of $\Csc(x)$ for $T>1/2$, both in
terms of the asymptotics ($\sim x^{\beta-2}$) and the full (and rather
simple) functional form\eq{k1_largeT}. Finally, our theoretical
analysis predicts how the decay exponent $\lam$ interpolates between
its value for $\k=0$ (standard \BM\ model) and the $\k=1$ limit of
slowly decorrelating observables, and reveals that the details of this
do depend on the choice of microscopic transition rates (Glauber
versus Metropolis).

One of the most surprising findings is the discontinuity in the
approach to $\k=1$ for $T>1/2$: strictly at $\k=1$ the correlation
function cannot decay and so $\Csc(x)\equiv 1$ and $\Msc(\ta)\equiv
1$. For $\k$ just below 1, on the other hand, we found that these
scaling functions 
do decay for finite values of their arguments, and approach nontrivial
limit forms for $\k\to 1$. At first sight this seems strange:
as explained above, $C(t,\tw)$ is the average of $\k^j$ over the
distribution $P_j(t,\tw)$ 
of the number of jumps between $\tw$ and $t$, and as $\k\to 1$ this
average must approach unity. Indeed, this statement is true for any
{\em fixed} pair of times $t$ and $\tw$. The discontinuity only arises
in the limit of large times, where $P_j$ consists of two essentially
separate parts. The first contains finite $j$, corresponding to histories where
the trap energies remain of the same order as the initial one: the
system wanders among deep traps. The contribution to the correlation
function arising from this part of $P_j$ is indeed continuous for
$\k\to 1$. However, $P_j$ also has a second part (containing a finite
fraction of its probability mass) at values of $j$ which are {\em
divergent} with $\tw$, presumably -- for consistency with the scaling
of the hopping rate -- as $\tw^{2-\beta}$. This part arises from
histories which make an excursion from deep traps to the top of
the landscape. In the limit $\tw\to\infty$,
the correlation decays to zero for any such history, as long as $\k<1$
(i.e.\ for {\em any} small but nonzero $1-\k$). This is why the
scaling functions, which are calculated in precisely this long-time
limit, are discontinuous at $\k=1$. For finite $\tw$ one expects to
see the crossover between correlation functions which essentially do
not decay, and the limiting forms calculated for $\k\to 1$, to occur
for $\k$-values with $\k^{\tw^{2-\beta}}$ of order one,
i.e. $1-\k\sim\tw^{\beta-2}$ for large $\tw$.

This argument also tells us how to interpret the $\k\to 1$ limit of
$\Csc(x)$: it is the probability of never having escaped to the
shallow traps at the top of the energy landscape. It is rather
intriguing to see that this can be calculated exactly while the $\k\to
0$ correlation, i.e.\ the probability of not having jumped at all,
cannot as far as we are aware.

Finally, it is interesting to see what happens to our naive estimate
from above, $C\approx\exp[(\ln\k)\int_{\tw}^t\!dt'\,\Gamma(t')]$, for
$T>1/2$. This would give $C(t,\tw)=\exp[-{\rm
const}\times(1-\k)(t^{2-\beta}-\tw^{2-\beta})]$ which decays on
subaging timescales $t-\tw\sim\tw^{\beta-1}$. Applied to the entire
correlation function this is plainly wrong: the correlation function
must decay more slowly as $\k$ increases, but for $\k=0$ we have
simple aging, so $\k>0$ must also give simple aging or an even slower
decay. We now see that 
the argument only applies to the decorrelation caused by large
(diverging with $\tw$) numbers of jumps, where it predicts correctly
that the resulting contribution to the correlation function vanishes
in the scaling limit. The remainder of the correlation function
remains finite (and has simple aging), however, because it relates to
the jumps among the deep traps which happen at rate $\sim 1/t$, rather
than to the total number of jumps which is dominated by the shallow
traps.

\section{Bouchaud model}
\label{sec:bouchaud}

The dynamics of the \BM\ model can be interpreted as being slowed down
by entropic barriers at low temperature: the Glauber (or Metropolis)
transition rates penalize large energy increases, so that the system
is forced to search for lower energy states of which there are an ever
decreasing number. To complete our analysis, we now compare with the
Bouchaud trap model~\cite{Bouchaud92}, where jumps take place with
rate $\exp(\beta E)$ {\em independently} of the energy of the arrival
trap; thus almost all jumps return the system to the top of the energy
landscape. Slow dynamics then arises purely from activation effects,
i.e.\ the decreasing rate with which jumps from increasingly deep
traps can take place. We will be interested to know in particular
whether slowly decorrelating observables again produce straight-line
FD plots.

Applying the intuitive reasoning from the previous section to the
Bouchaud model, we see that once the system has returned to the top of
the landscape it will make a large number of fast jumps, which will
decorrelate the values of the observable $m$ for any $\k<1$. Thus, the
correlation function $\Csc(x)$ should be completely independent of
$\k$ for $0\leq \k<1$ since, in the scaling limit, correlations are
maintained only if no jumps at all take place ($j=0$).

If the Bouchaud model, the natural prescription for the coupling to
the field is to modify the escape rate from a trap from $\exp(\beta
E)$ to $\exp[\beta(E-hm)]$~\cite{BouDea95,FieSol02}. This is just the
original rate multiplied by $\exp(-\beta h m)$, and one is led more
generally to consider an arbitrary trap model with multiplicative
rates~\cite{BouDea95,Ritort03,Sollich03}:
\be
w(E',m'\ar E,m) = e^{\beta h[(1-\zeta)m'-\zeta m]} w(E'\ar E)
\label{mult_rates}
\ee
The notation here should be self-explanatory; the arrow points from
the departure to the arrival trap. In the absence of a field all
governing equations are as before, except for the replacement of
$w(E'-E)$ by $w(E'\ar E)$. The master equation then reads
\be
\dd{t} P(E,t) = -\Gamma(E)P(E,t) + 
\int\! dE'\, \rho(E) w(E\ar E') P(E',m',t)
\label{master_mult}
\ee
with $\Gamma(E)=\int\! dE'\, \rho(E') w(E'\ar E)$, while the
magnetization decay function obeys
\be
\dd{t} M(t|E) = -\Gamma(E)M(t|E) + 
\k \int\! dE'\, M(t|E') \rho(E') w(E'\ar E)
\label{M_eq_mult}
\ee
and the two-time correlation function is given by\eq{corr} as before.
To get the response, one can start again from\eq{R_start}. The change
in $P(E,m,\tw+\dt)$ from its value without a field is given by the
direct analogue of\eq{dP},
\bea
\fl \D P(E,m,\tw+\dt) &=& \dt\,\int\!d\Ew\,d\mw \times
\nonumber\\
\fl & & \left[
 \rho(E)  \rho(m |\mw) \D w(E,m\ar\Ew,\mw) \rho(\mw) P(\Ew, \tw)
\right.
\nonumber\\
& &
\left.
-\rho(\Ew) \rho(\mw|m)  \D w(\Ew,\mw\ar E,m)  \rho(m)  P(E,  \tw)
\right]
\eea
Now to linear order in $h$, $\D
w(E,m\ar\Ew,\mw) = \beta h[(1-\zeta)m-\zeta \mw] w(E\ar
\Ew)$. Inserting this and integrating over $\mw$, and then over $m$
in\eq{R_start}, yields
\bea
\fl TR(t,\tw) &=&
\int\! dE\,d\Ew\,
M(t-\tw|E)
\left\{(-\zeta\k+1-\zeta)    \rho(E)  w(E\ar \Ew) P(\Ew,\tw)
\right.
\nonumber\\
\fl & &{}+{}\left.
                  [\zeta - (1-\zeta)\k] \rho(\Ew)w(\Ew\ar E) P(E,\tw)
\right\}
\label{R_mult1}
\eea
As in the case $\k=0$ one can now relate the response function to
derivatives of the correlation function. From\eq{corr} together
with\eq{M_eq_mult} one has
\bea
\fl \dd{t}C(t,\tw) &=& \int\! dE\, 
\left[\rule{0mm}{5mm}-\Gamma(E)M(t-\tw|E)\right.
\nonumber \\
\fl & & {}+{} \left.
\k \int\! dE'\, M(t-\tw|E') \rho(E') w(E'\ar E)\right]P(E,\tw)
\label{C_t}
\eea
For the derivative w.r.t.\ the earlier time $\tw$ one gets similarly,
by also using\eq{master_mult}, 
\be
\fl \dd{\tw}C(t,\tw) = (1-\k)\int\! dE\, dE'\, M(t-\tw|E') \rho(E')
w(E'\ar E) P(E,\tw)
\ee
This is proportional to the first term in the expression\eq{R_mult1}
for the response, while the second term in\eq{R_mult1} is proportional
to the first one in\eq{C_t}, bearing in mind the definition of
$\Gamma(E)$. Altogether one gets the simple relation
\be
TR(t,\tw) =
(1-\zeta)(1+\k)\dd{\tw}C(t,\tw)-[\zeta-(1-\zeta)\k]\dd{t}C(t,\tw)
\label{BD}
\ee
which reduces to the known
result~\cite{BouDea95,Ritort03,Sollich03} for $\k=0$ as it should and
applies to {\em all} trap models with multiplicatively perturbed
rates\eq{mult_rates}. At 
high temperatures, where a time-translation invariant equilibrium
state is reached, $\partial C/\partial\tw=-\partial C/\partial t$. As
required for consistency with equilibrium FDT, the coefficients of
these two quantities in\eq{BD} add up to unity. Out of equilibrium at
low temperatures, an equilibrium FD relation is still recovered when
the second coefficient vanishes, i.e.\ when $\zeta=\k/(1+\k)$. It is
easy to see that this is exactly the case where the exit rate
$\Gamma(E,m)$ from a given state is unperturbed by the field $h$ (to
linear order), just as for $\k=0$~\cite{Ritort03,Sollich03}. This
implies that for slowly decorrelating observables ($\k\to 1$) one only
gets a straight-line FD plot when 
$\zeta=1/2$, i.e.\ when the hopping rates depend on the difference
$m'-m$ as they do for Glauber dynamics.
%

For the Bouchaud model specifically, where $w(E'\to E) = \exp(\beta
E)$, equation\eq{M_eq_mult} for $M(t|E)$ is easily solved by LT to
give
\be
\fl \hat{M}(s|E) = \frac{1}{s+e^{\beta E}}\left(1+
\frac{\k e^{\beta E} \hat{G}(s)}{1-\k+\k s\hat{G}(s)}\right), \qquad
\hat{G}(s) = \int\!dE\,\frac{\rho(E)}{s+e^{\beta E}}
\ee
The scaling limit for $P(E,t)$ is reached for large $t$ and low $E$
with $t e^{\beta E}$ of $\order(1)$. This corresponds to small $s$, of
order $e^{\beta E}$. Since $\hat{G}(s)\sim s^{T-1}$ for small $s$, one
sees that in this regime the second term in $\hat{M}(s|E)$ always
becomes negligible compared to the first, as long as $\k<1$. This
means that the $\k$-dependence drops out, and $M(t|E)=\exp(-te^{\beta
E})$ in the scaling regime. The correlation function only picks up
these scaling contributions and is therefore $\k$-independent as
expected.

The same argument does not apply to the response, which from\eq{BD} is
obviously dependent on $\k$. The reason is clear from\eq{R_mult1}: the
contribution from $M(t-\tw|E)$ is weighted with extra factors of
either $\rho(E)$ or $w(\Ew\ar E)$ ($=e^{\beta E}$ in the Bouchaud
model). This means that the behaviour of $M(t-\tw|E)$ for
$E=\order(1)$ is dominant, and this does depend on $\k$.

\subsection{Arguments for straight-line FD plots}

The Bouchaud trap model provides a useful point of reference from
which to revisit the reasoning put forward in~\cite{JunBer04} for the
emergence of straight-line FD plots in the limit $\k\to 1$. This has
at its core the idea of considering first the change of the magnetization $m$
during $j$ jumps, and then to average over the distribution of the
number of jumps between $\tw$ and $t$. There is no problem with this
for the case of the correlation function, since in the absence of a
field the magnetization is just being ``convected along'' with the usual
trap model dynamics, without itself affecting the dynamics.

For the susceptibility, on the other hand, the situation is somewhat
more subtle. Consider a field switched to some nonzero value (and held
there) at time $\tw$. Then one can write generally
\bea
\chi &=& \dd{h} \sum_j \int\!dm\,d\mw\,d\Ew\, m P(m|j,h,\Ew,\mw,t-\tw)
\times
\nonumber\\
& & {}\times{} P_j(h,\Ew,\mw,t-\tw)P(\Ew,\mw,\tw)
\label{Ivan}
\eea
where $P(m|j,t-\tw,h,\Ew,\mw)$ is the distribution of $m$ at the end
of a time interval $t-\tw$ of constant field strength $h$, given that
$j$ jumps have taken place during this time and the system started from a
trap with energy $\Ew$ and magnetization $\mw$. $P_j(\ldots)$ is the
distribution of the number of jumps, which depends on the same
variables. Finally, $P(\Ew,\mw,\tw)$ is the distribution of trap
energies and magnetizations at $\tw$, which can be written as
$P(\Ew,\tw)\rho(\mw)$ if as we assume the field has been off up until time
$\tw$. Writing the $m$-average in\eq{Ivan} as $\bar{m}_j(h,\Ew,\mw,t-\tw)$ one
thus has
\be
\fl \chi = \dd{h} \sum_j \int\!d\Ew\,d\mw\, \bar{m}_j(h,\Ew,\mw,t-\tw)
P_j(h,\Ew,\mw,t-\tw)\rho(\mw)P(\Ew,\tw)
\label{Ivan2}
\ee
To complete the argument of \JB~\cite{JunBer04} two assumptions now
need to be made: 
(A) $\bar{m}_j(h,\Ew,\mw,t-\tw)$ is the same as the average of $m$
{\em directly after} the $j$-th jump starting from a state with $\Ew$
and $\mw$. (B)
The distribution of the number of jumps $P_j(\ldots)$ is independent
of $\mw$, so that $\bar{m}_j(\ldots)$ can be averaged separately over
$\mw$. The first assumption is in general invalid because the fact
that there has not been a further jump between the $j$-th jump and
time $t$ favours magnetizations $m$ of the same sign as the field
$h$. Indeed, in the standard ($\zeta=1$, $\k=0$) Bouchaud model
assumption (A) would imply that $\bar{m}_j=0$ for all $j\geq 1$ since
the distribution of $m$ after any jump is unbiased. But this cannot be
correct since then at long time differences, where the probability
that at least one jump has taken place approaches one, the
susceptibility $\chi$ 
would have to drop to zero. Assumption (B) is also in general not
correct; e.g.\ in the standard Bouchaud model the jump probabilities $P_j$
clearly depend on the value of the starting magnetization $\mw$ since
the distribution of the time until the first jump does.

There is, however, one scenario in which these objections do not
apply: if the exit rate $\Gamma(E,m)$ from a trap of energy $E$ and
magnetization $m$ is independent of $h$ (and hence of $m$), both
assumptions (A) and (B) are correct because the field biases neither
the probability of not having jumped since the $j$-th jump, nor the
probability of $j$ jumps having occurred. In the Bouchaud model, this
field-independence of the exit rate holds when $\zeta=\k/(1+\k)$. One
then works out easily, by considering how the Gaussian distribution of
$m$ changes with every jump, that
\be
\bar{m}_j(\ldots)=\k^j\mw+\beta
h(1-\zeta)(1+\k)(1-\k^j)
\label{bar_m_Bouchaud}
\ee
The susceptibility is then $\chi=\beta(1-\zeta)(1+\k)\langle
1-\k^j\rangle=\beta\langle 1-\k^j\rangle$, where the average is over
the distribution of jumps $\int d\Ew\,P_j(\Ew,t-\tw)P(\Ew,\tw)$. Since
the correlation function is $C=\langle \k^j\rangle$, it follows that
$T\chi=1-C$ and one has equilibrium FD behaviour. This is entirely
consistent with the exact relation\eq{BD}.

In the \BM\ model, assumption (A) can still be justified for $\k\to
1$: one can show~\cite{JunBer04} that the exit rates are affected by
the field via terms of $\order((1-\k)h)$, which vanish in the limit of
slow decorrelation. One is then allowed to find $\bar{m}_j$ from the
distribution of $m$ directly after the $j$-th jump, giving at low
temperatures $\bar{m}_j(\ldots) = \k^j\mw +
2h(1-\k^j)$~\cite{JunBer04}. However, for $\k\to 1$
one needs a number of jumps $j=\order(1/(1-\k))$ to see any
significant response of the system, and similarly any significant
decay of the correlation function. It is then not clear that
assumption (B) can still be justified, since the small $O(1-\k)$
changes of each exit rate accumulated over this many jumps could still
add up to a nontrivial $\mw$-dependence of $P_j(t,h,\Ew,\mw,\tw)$.

We note finally that, for $T>1/2$ and $\k\to 1$ in the \BM\ model, one
can state the argument for a straight-line FD plot without actually
requiring assumption (B): histories with a finite number of jumps $j$
do not contribute to $\chi$ because for $\k\to 1$ the magnetization
remains pinned to its initial value, whatever the field $h$. On the
other hand, histories with diverging $j$ always give the equilibrium
susceptilibity because they pass through the equilibrated shallow
traps at the top of the landscape. Because $C$ is just the probability
of $j$ being finite as argued in Sec.~\ref{sec:disc}, this gives the
equilibrium FD relation $\chi=\chi_{\rm eq}(1-C)=\beta(1-C)$.

\section{Conclusion}
\label{sec:conclusion}

We have studied trap models with observables that decorrelate slowly,
by a factor $\k\approx 1$ with each jump. Our motivation was to
clarify and extend with an analytical study the interesting
observations of \JB\ for a spin model which leads to trap model
dynamics of this type~\cite{JunBer04}. Our analysis showed that in the
limit of long times, correlation and response functions are determined
by two scaling functions: $\Psc(\om)$, which gives the scaling of the
distribution across trap energies, and $\Msc(\tau)$, which depends on
$\k$ and encapsulates the decay of the observable with
scaled time. Both of these are determined by integro-differential
equation which can be solved exactly at $T=0$ and by appropriate
numerical iteration procedures otherwise.

We focussed mostly on the \BM\ trap model which has entropic barriers
that slow its dynamics at low temperature. In Sec.~\ref{sec:kappa0} we
considered this first for fast decorrelation, $\k=0$, to extend previous
results which had been limited to $T=0$. The results already
reveal changes in the dynamics at $T=1/2$, i.e.\ half the glass transition
temperature $\Tg=1$ of the model: for $T<1/2$ the aging behaviour is
simple, with the average hopping rate scaling as the inverse of the
age, while for $T>1/2$ a power law scaling obtains which is dominated
by rare excursions of the system to the top of its energy landscape,
i.e.\ to the shallow traps. Correspondingly the initial slope of the
FD plot of susceptibility versus correlation is $1/\Teff$ with
$\Teff=1/2$ for $T<1/2$, and $1/T$ for larger temperatures. The
long-time susceptibility $\chi(t,\tw)$ for $t\gg\tw$, i.e.\ for
$C(t,\tw)\to 0$, remains constant throughout the glassy region $T<1$,
effectively ``freezing'' to its value at $T=1$ as temperature is
lowered through the glass transition.

In Sec.~\ref{sec:T0} we looked at the complementary case of $T=0$ but
slow decorrelation, $\k>0$. Here all scaling functions can be found
exactly. The initial slope of the FD plot, $1/\Teff=2$, remains
$\k$-independent, while the asymptotic (for $C\to 0$) slope $2\k$
grows linearly with $\k$. In the limit $\k\to 1$ of slow decorrelation
the two slopes coincide, and the FD plot becomes a straight line.

Finally, in Sec.~\ref{sec:kappa_T} we extended the analysis to general
decorrelation ($\k>0$) and nonzero temperature. Scaling functions now
need to be found numerically, but we were able to predict the
exponents for their power-law decays in closed form. Surprisingly,
these exponents depend on whether Glauber or Metropolis transition
rates are used, contrary to the usual expectation that such choices
only have minor quantitative effects. We were able to show explicitly
that the initial slopes of the FD plots are $\k$-independent, and that
in the limit $\k\to 1$ of slow decorrelation the FD plots become
straight lines as argued qualitatively by \JB~\cite{JunBer04}. Their
slopes, coinciding as they must with the $\k$-independent initial slopes, 
correspond to FD relations with an effective temperature $\Teff=1/2$
for $T<1/2$ and equilibrium FDT for $T>1/2$. We discussed in detail
the approach to the limit $\k\to 1$; for $T>1/2$ this
is discontinuous because rare excursions to the top of the landscape
achieve full decorrelation (and an associated equilibrium response)
whenever $\k$ is below unity by a nonzero amount $1-\k$.

We generalized to the Bouchaud trap model, and more broadly any trap
model with rates that are multiplicatively perturbed by the applied
field, in Sec.~\ref{sec:bouchaud}. Here it turns out that
straight-line FD plots always have an equilibrium slope of
$1/T$. However, they 
do not necessarily arise even in the limit $\k\to 1$, but
rely on a specific assignment of the field-dependence of the
transition rates which eliminates the effect of the field on the
residence time in a given trap. This example illustrated that simple
arguments for the existence of straight-line FD relations require some
caution because the implicit assumptions can be difficult to justify.

Looking at our results more broadly, it is intriguing to see that
slowly decorrelating observables {\em can} produce non-trivial
straight-line FD plots in trap model, something which can otherwise be
achieved only by choosing rates that violate detailed
balance~\cite{Ritort03}. The significance of this is that such plots
indicate that the effective temperature is (in the long-time,
non-equilibrium regime) independent of the pair of observation times;
this is one of the plausible requirements for $\Teff$ to be a
physically meaningful quantity. What remains unclear to us is why the
notional ``glass transition'' temperature which slowly decorrelating
observables seem to measure in trap models is {\em half} the usual
glass transition temperature $\Tg$ ($=1$ in our units). One way of
interpreting this result~\cite{JunBer04} is that slowly decorrelating
observables only pick up non-equilibrium effects caused by entropic
rather than energetic barriers. Alternatively, one could argue that
slowly decorrelating observables detect only persistently slow
dynamics. As soon as the system begins to exhibit intermittent
behaviour, where episodes of fast dynamics in the effectively
equilibrated parts of phase space alternate with periods of slow
wandering around regions of low energy, effective equilibrium FD
relations are recovered. Slowly decorrelating observables can thus
help us to single out parts of the dynamics which can be meaningfully
associated with an effective temperature, but by the same token can be
blind to other aspects of the dynamics that remain clearly out of
equilibrium. How this trade-off operates in other glassy systems is
certainly worthy of further study. One challenge will be to integrate
finite spatial dimensionality into the picture. One expects, for
example, that the dynamics will then effectively decompose into that
of independent subsystems whose finite size is set by a time-dependent
correlation length~\cite{HeuDolSak05}. Decorrelation can then not be
arbitrarily slow: if each subsystem contains $N_{\rm eff}$ degrees of
freedom, say, one would expect $\k<1-\order(1/N_{\rm eff})$. The limit
of slow decorrelation we focussed on here would then require
relatively large correlated subsystems.

{\bf Acknowledgements:} We are grateful to Eric Bertin, Ivan Junier
and Felix Ritort for helpful and inspiring discussions.

\section*{References}

\bibliography{/home/psollich/references/references}
\bibliographystyle{unsrt}

\end{document}